\documentclass[conference]{IEEEtran}
\usepackage{cite}
\usepackage{amsmath,amssymb,amsfonts,amssymb}
\usepackage{algorithmic}
\usepackage{graphicx}
\usepackage{textcomp}
\usepackage{xcolor}
\usepackage{tikz}
\usepackage{url}
\usepackage{multirow}
\usepackage[notextcomp]{stix}

\usepackage[hidelinks]{hyperref}

\usepackage{soul}
\soulregister\cite 7

\def\BibTeX{{\rm B\kern-.05em{\sc i\kern-.025em b}\kern-.08em
    T\kern-.1667em\lower.7ex\hbox{E}\kern-.125emX}}

\definecolor{revcol}{rgb}{0 0 0}

\newcommand{\rev}[1]{\textcolor{revcol}{#1}}

\begin{document}

%\title{Your Eyes Tell: On the Limits of \\Textual Screen Peeking via
\title{Private Eye: On the Limits of \\Textual Screen Peeking via Eyeglass Reflections \\in Video Conferencing
}
\author{
\IEEEauthorblockN{Yan Long$^{\ast}$, Chen Yan$^{\dagger}$, Shilin Xiao$^{\dagger}$, Shivan Prasad$^{\ast}$, Wenyuan Xu$^{\dagger}$, and Kevin Fu$^{\ast}$}
\IEEEauthorblockA{$^{\ast}$Electrical Engineering and Computer Science, University of Michigan, Ann Arbor, USA \\
$^{\dagger}$ College of Electrical Engineering, Zhejiang University, Hangzhou, China \\
\{yanlong, shprasad, kevinfu\}@umich.edu, \{yanchen, xshilin, wyxu\}@zju.edu.cn}
}
% \author{
% \IEEEauthorblockN{Yan Long$^{\ast}$, Chen Yan$^{\dagger}$, Shilin Xiao$^{\dagger}$, Shivan Prasad$^{\ast}$, Wenyuan Xu$^{\dagger}$, and Kevin Fu$^{\ast}$}
% \IEEEauthorblockA{$^{\ast}$Electrical Engineering and Computer Science, University of Michigan, Ann Arbor, USA \\
% $^{\dagger}$ College of Electrical Engineering, Zhejiang University, Hangzhou, China \\
% \{yanlong, shprasad, kevinfu\}@umich.edu, \{yanchen, xshilin,  wyxu\}@zju.edu.cn}
% }

\maketitle

\begin{abstract}
Personal video conferencing has become a new norm after COVID-19 caused a seismic shift from in-person meetings and phone calls to video conferencing for daily communications and sensitive business. Video leaks participants' on-screen information because eyeglasses and other reflective objects unwittingly expose partial screen contents.  Using mathematical modeling and human subjects experiments, this research explores the extent to which emerging webcams might leak recognizable textual \rev{and graphical} information gleaming from eyeglass reflections captured by webcams.  The primary goal of our work is to measure, compute, and predict the factors, limits, and thresholds of recognizability as webcam technology evolves in the future. Our work explores and characterizes the viable threat models based on optical attacks using multi-frame super resolution techniques on sequences of video frames. Our models and experimental results in a controlled lab setting  show it is possible to reconstruct and recognize \rev{with over 75\% accuracy}  on-screen texts that have heights as small as 10~mm with a 720p~webcam. We further apply this threat model to web textual contents with varying attacker capabilities to find thresholds at which text becomes recognizable. \rev{Our user study with 20~participants suggests present-day 720p webcams are sufficient for adversaries to reconstruct textual content on big-font websites. Our models further show that the evolution towards 4K~cameras will tip the threshold of text leakage to reconstruction of most header texts on popular websites. Besides textual targets, a case study on recognizing a closed-world dataset of Alexa top 100 websites with 720p webcams shows a maximum recognition accuracy of 94\% with 10 participants even without using machine-learning models. Our research proposes  near-term mitigations including a software prototype that users can use to blur the eyeglass areas of their video streams. For possible long-term defenses, we advocate an individual reflection testing procedure to assess threats under various settings,} and justify the importance of following the principle of least privilege for privacy-sensitive scenarios. 

\end{abstract}

\section{Introduction}

Online video calls have become ubiquitous as a remote communication method, especially since the recent COVID-19 pandemic that caused almost universal work-from-home policies in major countries~\cite{deng2020running,bick2020work,arntz2020working} and made video conference a norm for companies and schools to accommodate interpersonal communications even after the pandemic~\cite{CNBC, karl2022virtual, sabra2020zoom,norm}. 

While video conferencing provides people with the convenience and immersion of visual interactions, it unwittingly reveals sensitive textual information that could be exploited by a malicious party acting as a participant. Each video participant's screen could contain private information. The participant's own webcam could capture this information when it is reflected by the participant's eyeglasses and unwittingly provide the information to the adversary (Figure 1). We refer to this attack as a \textit{webcam peeking attack}. Furthermore, adversary capabilities will only continue to increase with improvements to resolution, frame rate, and more. It is thus important to understand the consequences and limits of webcam peeking attacks in present-day and possible future settings. 

\begin{figure}[!t]
	\centering
	\includegraphics[width=.4\textwidth]{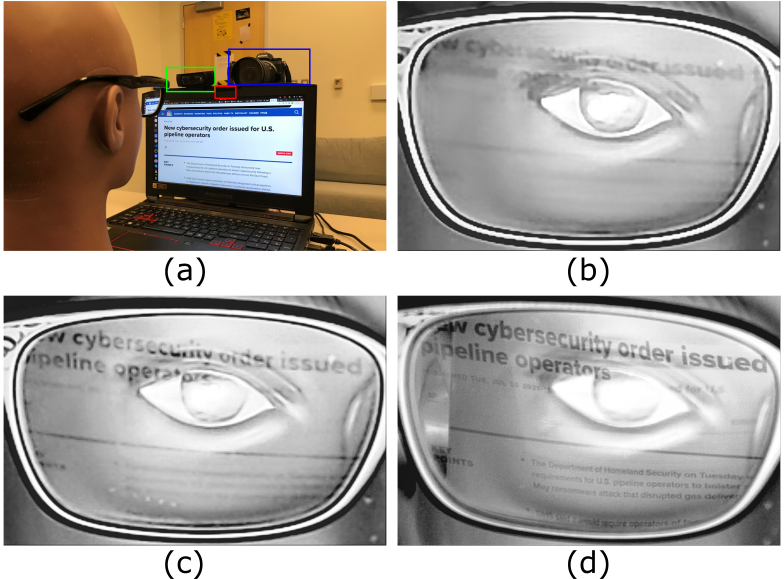}
		\vspace{-.1in}

	\caption{The optical emanations of the victim's screen are reflected by eyeglasses, captured by the victim's webcam, and streamed to the adversary, which can then be used to reconstruct the screen contents. The experimental setup (a) with a laptop built-in webcam (b) (red box, 720p), an external Logitech webcam (c) (green box, 1080p), and a Nikon DSLR (d) (blue box, 4K) helps us predict the future fidelity of the attacks as  video conferencing technologies evolve.} 
	% \vspace{-.2in}
	\label{fig:threat_model}
\end{figure}
% \vspace{-.2in}

{\let\thefootnote\relax\footnote{{\href{https://doi.ieeecomputersociety.org/10.1109/SP46215.2023.00050}{\textcolor{blue}{This paper is officially published in the IEEE SP'23 Proceedings}}}}}

Previous work shows that similar attacks exploiting optical reflection off nearby objects in controlled setups are feasible, such as observing teapots on a desk with high-end digital single-lens reflex (DSLR) cameras and telescopes at a distance~\cite{backes2008compromising,backes2009tempest}. The challenge and characterization of peeking using the more ubiquitous webcams, however, are qualitatively different due to the lower-quality images of present-day webcams. The lower-quality webcam images are caused by unique types of distortions, namely the shot and ISO noise due to insufficient light reception, and call for new image-enhancing techniques. In addition, new mathematical models and analysis frameworks are needed to understand the threat model of webcam peeking attacks. Finally, this new threat model requires a dedicated evaluation to clarify the potential threats and mitigations to the average video conference user.

There are many types of media that can leak over optical reflections, including text and graphics.  We focus on textual leakage in this work as it’s a natural starting point for measurable recognizability and modeling of the fundamental baseline of information leakage, \rev{but also provides insights into the leakage of non-textual information such as inferring displayed websites through recognizing graphical contents on the screen.} We seek to answer the following three major questions: $\mathcal{Q}1$: What are the primary factors affecting the capability of the webcam peeking adversary? $\mathcal{Q}2$: What are the physical limits of the adversary's capability in the present day and the predictable future, and how can adversaries possibly extend the limits? $\mathcal{Q}3$: What are the corresponding threats of webcam peeking against cyberspace targets and the possible mitigations against the threats?

To answer $\mathcal{Q}1$, we propose a simplified yet reasonably accurate mathematical model for reflection pixel size. The model includes factors such as camera resolution and glass-screen distance and enables the prediction of webcam peeking limits as camera and video technology evolve. By using the complex-wavelet structural similarity index as an objective metric for reflection recognizability, we also provide semi-quantitative analysis for other physical factors including environmental light intensity that affect the signal-to-noise ratio of reflections. 

To answer $\mathcal{Q}2$, we analyze the distortions in the webcam images and propose multi-frame super resolution reconstruction for effective image enhancement to extend the limits. We then gather eyeglass reflection data in optimized lab environments and evaluate the recognizability limits of the reflections  through both crowdsourcing workers on Amazon Mechanical Turk and optical character recognition models. The evaluation shows over 75\% accuracy on recognizing texts that have a physical height of 10~mm with  a 720p~webcam. 

To answer $\mathcal{Q}3$, we focus on web textual targets to  build a benchmark that enables meaningful comparisons between present-day and future webcam peeking threats. We first map the limits derived from the model and evaluations to web textual content by surveying previous reports on web text size and manually inspecting fonts in 117~big-font websites. Then, we conduct a user study with 20~participants and play a challenge-response game where one author acts as an adversary to infer HTML contents created by other authors. Results of the user study suggest that present-day 720p~webcams can peek texts in the 117~big-font websites and future 4K~webcams are predicted to pose threats to header texts from popular websites. We investigated the underlying factors enabling easier webcam peeking in the user study by analyzing the correlation between adversary recognition accuracy and multiple factors. We found, for example, user-specific parameters including browser zoom ratio play a more important role than the glass-screen distance. Besides texts, we also explored the feasibility of recognizing websites through graphical content with 10 participants and observed accuracies as high as 94\% on recognizing a closed-world dataset of Alexa top 100 websites.

Finally, we discuss possible  near-term mitigations including adjusting environmental lighting and blurring the glass area in software. We also envision long-term solutions following an individual reflection assessment procedure and a principle of least privilege. In summary, the goal of this work is to provide a theoretical foundation and benchmark for the study of emerging webcam peeking threats with evolving webcam technologies and the development of securer video conferencing infrastructures. We summarize our main contributions:

\begin{itemize}
    \item \rev{Our work quantifies the limits and primary factors that predict the degree of information leakage from webcam peeking by using theoretical modeling and experimentation. This characterization helps predict future unknown vulnerabilities tied to the limits of evolving webcam technologies that do not yet exist.}

    \item \rev{A benchmark centering on web textual targets that enables comparisons of webcam peeking threats. Our benchmarking methodology builds upon web text design conventions and a 20-participant user study on present-day cameras  such that the benchmark can be applied to both hypothetical and emerging cameras in the coming years.}

    \item \rev{Analysis on near-term mitigations including using software-based blurring filters and changing physical setups as well as  possible long-term defenses by proactive testing and following a principle of least privilege. Our analysis investigates the potential effectiveness and  implementation methods of different protections. }
\end{itemize}

\section{Threat Model \& Background}

\subsection{Threat Model}
In this work, we study the webcam peeking attack during online video conferences, where the adversary and the victim are both participants. We assume the device the victim uses to join the video conference consists of a display screen and either a built-in or an external webcam that is mounted on the top of the screen as in most cases, and the victims wear glasses with a reflectance larger than 0, i.e., at least a portion of the light emanated by the monitor screen can be reflected from the glasses to the webcams. We do not enforce constraints on the devices used by the adversary. When the adversary launches the attack, we assume the victim is facing the screen and webcam in the way that the screen emanated light has a single-reflection optical path into the webcam through the eyeglass lens's outer surface. We do not assume the adversary has any control or information on the victim's device.

We assume that the victim's up-link video stream is enabled during the attack, and the adversary can acquire the down-link video stream of the victim. The adversary can achieve that by either directly intercepting the down-link video stream data, or recording the victim's video with the video conferencing platform being used or even third-party screen recording services. Since the webcam peeking attack does not require active interaction between the victim and the adversary, the adversary does not need to attempt a real-time attack but can store the video recording and analyze the videos offline. 

\subsection{Glasses} \label{sec:glasses}
The most common types of glasses that people wear in a video conferencing setting are prescription glasses~\cite{holden2016global} and blue-light blocking (BLB) glasses~\cite{palavets2019blue,blb_report}. BLB glasses can either have prescriptions with BLB coating or be  non-prescription (flat). The reflectance and curvature of glass lenses are the two most important characteristics in the process of reflecting screen optical emanations.  

\textbf{Reflectance.} Reflectance of a lens surface is the ratio between the light energy reflected and the total energy incident on a surface\cite{schott}. Reflectance is wavelength-dependent. The higher the reflectance, the more light can  be reflected to  and captured by a webcam. 

% Prescription glasses normally have a reflectance higher than 10\% in the visible light spectrum with the highest reflectance being in the green and blue-light range~\cite{ozdemir2016evaluation,schott}. Previous researches show that commercial BLB glasses often have about 50\% reflectance in the blue-light spectrum~\cite{leung2017blue,carlson2019comparison}. 

\textbf{Curvature.} Curvature of a lens surface represents how much it deviates from a plane. The concepts of curvature, radius, and focal length of an eyeglass lens are used interchangeably on different occasions and are related by: 
$
Curvature = 1/Radius = 2/Focal Length
$. Smaller curvature  leads to larger-size reflections. Both the outer and inner surfaces of a lens can reflect, but the outer surface often has smaller curvature and thus produce better quality reflections (Appendix~\ref{apdx:dev}). This paper refers to the eyeglass lens curvature/radius/focal length as that of the outer surface if not specified otherwise.

% Blue-light blocking (BLB) glasses are made of or blue-light filtering spectacle lenses which can  prevent (a portion of) blue light energy coming through the lenses and reaching human eyes. Blue light is a type of electromagnetic radiation with a wavelength in the range of around 400-500nm. As indicated by its name, these radiations are perceived by human eyes in colors similar to blue, which are of the shortest wavelengths in the visible light spectrum (around 400-780nm) and thus carry the highest energy of visible-light radiation. Blue light has been drawing increasing attentions because of the hypothesis and preliminary verification on its potential to induce photo-chemical damages to retina~\cite{ham1976retinal,ham1978histologic,ham1980nature},  affect the sleep quality of humans~\cite{kessel2011sleep,kimberly2009amber}, and cause fast eye fatigues~\cite{palavets2019blue}. etc. While there are ongoing debates on whether reducing blue light can effectively reduce the negative effects~\cite{leung2017blue,landers2009effect}, BLB glasses have gained increasing popularity in the global general population~\cite{palavets2019blue}. Report shows that the global market of BLB glasses is estimated to grow from USD 22 million in 2019 to USD 22 million by 2025(, suggesting dozens of millions of potential users.) 

\subsection{Digital Camera Imaging System}
Digital cameras have sensing units uniformly distributed on the sensor plane, each of which is a Charge-coupled Device (CCD) or Complementary Metal-oxide-semiconductor (CMOS) circuit unit that converts the energy of the photons it receives within a certain period of time, i.e., the exposure time, to an amplitude-modulated electric signal. Each sensing unit then corresponds to a ``pixel'' in the digital domain.  The quality of a digital image to human perception is mainly determined by its pixel resolution, color representation, the amount of received light that is of our interest, and various imaging noise. The  2 key imaging parameters that are closely related to webcam peeking attacks are described below.  

\textbf{Exposure Time.} Theoretically, the longer the exposure time, the more photons will hit the imaging sensors, and thus there can be potentially more light of interest captured. The images with a longer exposure time will generally be brighter. The downside of having a longer exposure time is the aggravated motion blur when imaging a moving object.

\textbf{ISO Value.} The ISO value represents the amplification factor of the photon-induced electrical signals.  In darker conditions, the user can often make the images brighter by increasing the ISO value. The downside of having a higher ISO is the simultaneous amplification of various imaging noises.

\subsection{Text Size Representations} \label{sec:txtreps}

It is important to select proper representations of text size in both digital and physical domains since the size of the smallest recognizable texts is the key metric for webcam peeking limits. When texts are digital, i.e., in the victim's software such as browsers and in the webcam image acquired by the adversary, we use point size and pixel size to represent the text size respectively. In the physical domain, i.e., when the texts are displayed on users' screens as physical objects, we use the cap height of the fonts and the physical unit mm to represent the size as it is invariant across different computer displays and enable quantitative analysis of the threats. Cap height is the uniform height of capitalized letters when font style and size are specified and is thus  usually used as a convenient representation of physical text size and the base for other font parameters~\cite{arditi2004adjustable,arditi2005serifs}.

\section{Webcam Peeking Through Glasses}\label{sec:webcampeeking}

In this section, we start with a feasibility test that reveals the 3 key building blocks of the webcam peeking threat model, namely (1) reflection pixel size, (2) viewing angle, and (3) light signal-to-noise ratio (SNR). For the first two building blocks, we develop a mathematical model that quantifies the related impact factors. For light SNR, we analyze one major factor it encompasses, i.e., image distortions caused by shot noise, and investigate using multi-frame super resolution (MFSR) to enhance reflection images. We will analyze other physical factors that affect light SNR in Section \ref{sec:factors}.  Experiments are conducted with the Acer laptop with its built-in 720p webcam, the pair of BLB glasses, and  the pair of prescription glasses described in Appendix \ref{apdx:dev}.

% \begin{figure}[!t]
% 	\centering
% 	\includegraphics[width=.4\textwidth]{figs/template_match.pdf}
% 	\caption{\ly{To use bar graphs} The number of correctly recognized letters with different sizes (heights) in the feasibility test. 10 different frame positions generate different recognition performance.} 
% 	\label{fig:recog}
% \end{figure}

\begin{figure}[!t]
	\centering
	\includegraphics[width=.45\textwidth]{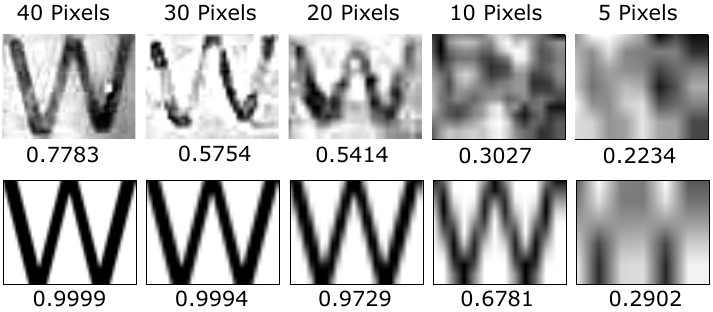}
    % \vspace{-.1in}
	\caption{(Upper) The captured images of the reflections. Compared with the ideal reflections, additional distortions exist that undermine image recognizability. (Lower) The estimated ideal reflections in the feasibility test corresponding to letters with a height of 80, 60, 40, 20, 10 mm respectively. The images are subjected to aliasing when enlarged.} 
	% \vspace{-.2in}
	\label{fig:template_sizes}
\end{figure}

\subsection{Feasibility Test}
We conduct a feasibility test of recognizing single alphabet letters with a similar setup as in Figure \ref{fig:threat_model}. A mannequin wears the BLB glasses with a glass-screen distance of 30 cm. Capital letters with different cap heights (80, 60, 40, 20, 10 mm) are displayed and captured by the webcam. Figure \ref{fig:template_sizes} (upper) shows the captured reflections. We find that the 5 different cap heights resulted in letters with heights of 40, 30, 20, 10, and 5 pixels in the captured images. As expected, texts represented by fewer pixels are harder to recognize. The reflection pixel size acquired by adversaries is thus one key building block of the characteristics of webcam peeking attack that we need to model. In addition, Figure \ref{fig:template_sizes} (lower) shows the ideal reflections with these pixel sizes by resampling the template image. Comparing the two, we notice small-size texts are subjected to additional distortions besides the issue of small pixel resolution and noise caused by the face background, resulting in a bad signal-to-noise ratio (SNR) of the textual signals. 

To quantify the differences using objective metrics, we embody the notion of reflection quality in the similarity between the reflected texts and the original templates. We compared multiple widely-used image structural and textural similarity indexes including structural similarity Index (SSIM) \cite{wang2004image}, complex-wavelet SSIM (CWSSIM) \cite{sampat2009complex}, feature similarity (FSIM) \cite{zhang2011fsim}, deep image structure and texture similarity (DISTS) \cite{ding2020image} as well as self-built indexes based on scale-invariant feature transform (SIFT) features \cite{lindeberg2012scale}. Overall, we found CWSSIM which spans the interval $[0, 1]$ with larger numbers representing higher reflection quality produces the best match with human perception results.
Figure \ref{fig:template_sizes} shows the CWSSIM scores under each image. 

The differences show that the SNR of reflected light corresponding to the textual targets is another key building block we need to characterize. Finally, we notice that when we rotate the mannequin with an angle exceeding a certain threshold, the webcam images do not contain the displayed letters on the screen anymore. It suggests that the viewing angle is another critical building block of the webcam peeking threat model which acts as an on/off function for successful recognition of screen contents. In the following sections, we seek to characterize these three building blocks.

\begin{table}[!t]
    \centering
        \caption{Parameters for modeling reflection pixel size}
        % \vspace{-.1in}
    \begin{tabular}{|p{1.3cm} | p{6.5cm}| }
    \hline
    \textbf{Notation} &  \textbf{Parameter}\\ \hline
    $\;\;\;\;\;\;h_o$ &  Physical size (cap height) of the object on the screen \\ \hline
    $\;\;\;\;\;\;h_s$ & Physical size of the object's projection on the sensor \\ \hline 
    $\;\;\;\;\;\;s_p$ & Pixel size of the imaged object \\ \hline
    $\;\;\;\;\;\;h_i$ & Physical size of the object's virtual image \\ \hline
    $\;\;\;\;\;\;P$ & Physical size of a single imaging sensor pixel \\ \hline
    $\;\;\;\;\;\;N$ & Number of pixels the camera has in the dimension \\ \hline
    $\;\;\;\;\;\;W$ & Physical size of the imaging sensor in the dimension\\ \hline
    $\;\;\;\;\;\;f$ & Camera focal length \\ \hline
    $\;\;\;\;\;\;d_o$ &Distance between screen and glasses \\ \hline
    $\;\;\;\;\;\;d_i$  & Distance between glasses and virtual image \\ \hline
    $\;\;\;\;\;\;f_g$  & Focal length of the glasses convex outer surface \\ \hline
    
\end{tabular}
% \vspace{-.2in}
    \label{tab:imaging-param}
\end{table}

\subsection{Reflection Pixel Size} \label{sec:model}
In the attack, the embodiment of textual targets undergoes a 2-stage conversion process: digital (victim software) $\rightarrow$ physical (victim screen) $\rightarrow$ digital (adversary camera). In the first stage, texts specified usually in  point size in software by the user or web designers are rendered on the victim screen with corresponding physical cap heights. In the second stage, the on-screen texts get reflected by the glass, captured by the camera, digitized, and transferred to the adversary's software as an image with certain pixel sizes. Generally, more usable pixels representing the texts enable adversaries to recognize texts more easily. The key is thus to understand the mechanism of point size $\rightarrow$ cap height $\rightarrow$ pixel size conversion.

\textbf{Point Size $\rightarrow$ Cap Height.} Mapping between digital point size and physical cap height is not unique but dependent on user-specific factors and software. The conversion formula for most web browsers can be summarized as follows: 
\begin{equation} \label{eq:cap}
    h_o = \frac{4}{3}p_t \cdot \frac{H_{scr}}{N_{os}} \cdot s_{os} \cdot s_b  \cdot r_{cap}
\end{equation}
where $h_o$ is the physical cap height of the text, $\frac{4}{3}p_t$ is the number of display hardware pixels most web browsers use to render the text given a point size $p_t$, $H_{scr}$ is the physical height of the screen,  $N_{os}$ is the screen resolution on the height dimension set in the OS which can be equal to or smaller than the maximum supported resolution, $s_{os}$ and $s_b$ are the OS and browser zoom/scaling ratios respectively, and $r_{cap}$ is the ratio between the cap height and the physical point size which is on average $\frac{2}{3}$~\cite{arditi2004adjustable,arditi2005serifs}.

\begin{figure}[!t]
	\centering
	\includegraphics[width=.4\textwidth]{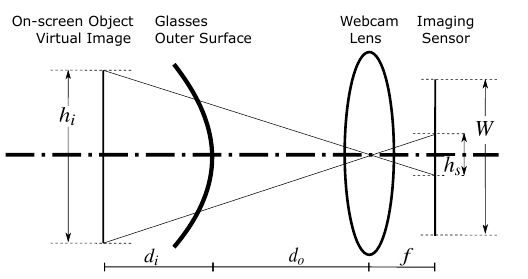}
	% \vspace{-.1in}
	\caption{The model of reflection pixel size. To better depict the objects, the sizes are not drawn up to scale. The screen overlaps with the webcam lens and is omitted in the figure.} 
	% \vspace{-.2in}
	\label{fig:imaging}
\end{figure}

\textbf{Cap Height $\rightarrow$ Pixel Size.} We would like to remind the readers that we only use pixel size to represent the size of texts living in the images acquired by the adversary\footnote{Since web/software designers sometimes also directly specify text size in pixel size ($\frac{4}{3}P_t$ in Equation \ref{eq:cap}), the two pixel sizes can be easily confused without explanation.}.  Figure \ref{fig:imaging} shows the model for  this conversion process. To simplify the model, we assume the glasses lens, screen contents, and webcam are aligned on the same line with the same angle. The result of this approximation is the loss of projective transformation information, which only causes small inaccuracies for reflection pixel size estimation in most webcam peeking scenarios. Figure \ref{fig:imaging} only depicts one dimension out of the horizontal and vertical dimensions of the optical system but can be used for both dimensions. In this work we focus on the vertical dimension for analysis, i.e., the reflection pixel size we discuss is the height of the captured reflections in pixels. We summarize the parameters of this optical imaging system model in Table \ref{tab:imaging-param}. Through trigonometry, we know 
% $$
% \begin{cases} 
% \frac{h_s}{f} = \frac{h_i}{d_o+d_i} \\ 
% h_s = s_pP \:\:\:\:\:\: \Rightarrow s_p = \frac{h_i}{d_o+d_i}\cdot \frac{f}{W} \cdot N\\
% P = \frac{W}{N}
% \end{cases}
% $$
% \begin{equation} \label{eq:reflect}
% \begin{split}
% \Rightarrow s_p = \frac{h_i}{d_o+d_i}\cdot \frac{f}{W} \cdot N
% \end{split}
% \end{equation}

\begin{equation} \label{eq:reflect}
\begin{cases} 
\frac{h_s}{f} = \frac{h_i}{d_o+d_i} \\ 
h_s = s_pP \:\:\:\:\:\: \Rightarrow \:\:\: s_p = \frac{h_i}{d_o+d_i}\cdot \frac{f}{W} \cdot N\\
P = \frac{W}{N}
\end{cases}
\end{equation}

As pointed out in Section \ref{sec:glasses}, the reflective outer surface of glasses is mostly convex mirrors which shrink the size of the imaginary object $h_i$ and also decrease $d_i$ compared to an ideal flat mirror. To calculate the reflection pixel size $s_p$ in this case, we can use the convex mirror equations~\cite{higashiyama2001perceived}
$$
\begin{cases} 
\frac{1}{(-f_g)} = \frac{1}{d_o} + \frac{1}{(-d_i)} \\
\frac{h_i}{h_o} = \frac{d_i}{d_o}
\end{cases}
$$
where $f_g$ is the focal length of the convex mirror which is half of the radius of the glasses lens and is defined to be positive. Plugging the above equations into Equation \ref{eq:reflect} we 
can then get 
\begin{equation}\label{eq:convex}
    s_p = \frac{ h_o f_g}{d_o^2 + 2d_o f_g} \cdot \frac{f}{W} \cdot N,
\end{equation}

% If we consider an ideal flat mirror, i.e., a glasses lens with a curvature of $0$ and thus $f_g = +\infty$, Equation \ref{eq:convex} degrades to 
% \begin{equation}\label{eq:flat}
% s_p = \frac{h_o}{2d_o}\cdot \frac{f}{W} \cdot N
% \end{equation}
% Comparing Equation \ref{eq:convex} with Equation \ref{eq:flat}, we can clearly see that increased curvature (decreased $f_g$) of the glasses lead to smaller pixel size of the reflections in the webcams.

The term $\frac{f}{W}$ of typical laptop webcams can be estimated to be in the range $1.1-1.4$ (see Appendix \ref{apdx:dev}). With the Acer laptop and BLB glasses and a glass-screen distance of $d_o=30$ cm, the estimated vertical pixel size of a $20$ mm-tall object displayed on the screen is in the range of $9.2-11.7$ pixels, which matches with the 10 pixels found in the feasibility test and verifies the accuracy of the model despite the approximation we made.

\begin{table}[!t]
    \centering
        \caption{The predicted feasible attack ranges for the viewing angle.}
        % \vspace{-.1in}
    \begin{tabular}{| p{3.5cm} | p{1.7cm}  | p{1.9cm}| }
    \hline
    \textbf{Type} &  \textbf{Theoretical} & \textbf{Measurement}\\ \hline
    Pres: All Page + Horizontal &  $\pm 15^{\circ}$ & $\pm 17^{\circ}$  \\ \hline
    Pres: Center + Horizontal &  $\pm 5^{\circ}$ & $\pm 8^{\circ}$\\  \hline
    Pres: All Page + Vertical & $\pm 9^{\circ}$& $\pm 13^{\circ}$\\ \hline
    Pres: Center + Vertical & $\pm 3^{\circ}$& $\pm 5^{\circ}$\\ \hline
    
    BLB: All Page + Horizontal &  $\pm 20^{\circ}$ & $\pm 25^{\circ}$  \\ \hline
    BLB: Center + Horizontal &  $\pm 10^{\circ}$ & $\pm 13^{\circ}$  \\ \hline
    BLB: All Page + Vertical &  $\pm 14^{\circ}$ & $\pm 19^{\circ}$  \\ \hline
    BLB: Center + Vertical &  $\pm 8^{\circ}$ & $\pm 10^{\circ}$  \\ \hline

\end{tabular}
% \vspace{-.2in}
    \label{tab:angle}
\end{table}

\subsection{Viewing Angle}
To model the effect of viewing angle and the range of angles that enables webcam peeking attack, we model the lens as spherical with a radius $2f_g$. A detailed derivation of the viewing angle model can be found in Appendix \ref{apdx:angle}. We consider two cases of successful peeking with a rotation of the glass lens. The first case All Page claims success as long as there exists a point on the screen whose emitted light ray can reach the camera. The second case Center claims success only if the contents at the center of the screen  have emitted lights that can be reflected to camera. Table \ref{tab:angle} summarizes the calculated theoretical angle ranges and the measured values. Both the theoretical model and measurements show that the webcam peeking attack is relatively robust to human positioning with different head viewing angles, which is validated later by the user study results (Section \ref{sec:user}).

\subsection{Image Distortion Characterization} \label{sec:distortion}
% Compared to reflection pixel size and viewing angle, factors affecting the light SNR of reflections is harder to fully quantify as it requires accurate optical models for the imaging sensors which are device-dependent. We thus aim to offer semi-quantitative analysis and start by focusing on the image distortion factor, since understanding the source of image distortions provides us with insights into the ways an adversary may improve the reflection quality and thus extend the limits of webcam peeking. We will study the  other environmental factors in Section \ref{sec:factors}.

Generally, the possible distortions are composed of imaging systems' inherent distortions and other external distortions. Inherent distortions mainly include out-of-focus blur and various imaging noises introduced by non-ideal camera circuits. Such inherent distortions exist in camera outputs even when no user interacts with the camera. External distortions, on the other hand, mainly include factors like motion blur caused by the movement of active webcam users.

\color{revcol}
\textbf{User Movement-caused Motion Blur.} When users move in front of their webcams, reflections from their glasses move accordingly which can cause blurs in the camera images. User motions can be decomposed into two components, namely  involuntary periodic small-amplitude tremors that are always present~\cite{elble2017tremor}, and  intentional non-periodic large-amplitude  movements that are occasionally caused by random events such as a user moving its head to look aside. By approximating user motions as displacements of $h_o$ and utilizing Equation~\ref{eq:convex}, the number of blurred pixels $\delta_p$ can be estimated by\footnote{We mainly consider motions that are parallel to the screen because generally, they  cause  larger blurs than other types of motions}: 

$$
\delta_p = \frac{ \delta^{T} f_g}{d_o^2 + 2d_o f_g} \cdot \frac{f}{W} \cdot N
$$
where $\delta^{T}$ is the motion  displacement amplitude within the exposure time of a frame.

For tremor-based motion, existing research suggests the mean displacement amplitude of dystonia patients’ head tremors is under 4 mm with a maximum frequency of about 6 Hz~\cite{elble2017assessment}. Since dystonia patients have  stronger tremors than healthy people, this provides an estimation of the tremor amplitude upper bound. With the example glass in Section~\ref{sec:model} and a 30 fps camera, the estimated pixel blur is under 1 pixel. Such a motion blur is likely to affect the recognition of extremely small reflections. Intentional motion is not a focus of this work due to its random, occasional, and individual-specific characteristics. We will experimentally involve the impacts of intentional user motions in the user study by letting users behave normally.

\color{black}

\textbf{Distortion Analysis.} To observe and analyze the dominant types of distortions, we recorded videos with the laptop webcam and a Nikon Z7 DSLR~\cite{z7} representing a higher-quality imaging system. The setup is the same as the feasibility test except that  \rev{we tested with both the still mannequin and a human to analyze the effects of human tremor}.  Figure \ref{fig:nnonuniform} (a) shows the comparison between the ideal reflection capture and the actual captures in three consecutive video frames of the webcam (1st row) and Nikon Z7 (2nd row) \rev{when the human wears the glasses}.  \rev{Empirically, we observed the following three key features of the video frames in this setup with both the mannequin and human} (see Appendix~\ref{apdx:distortion} for details): 
\begin{itemize}
    \item Out-of-focus blur and \rev{tremor-caused motion blur}   are generally negligible \rev{when the reflected texts are recognizable}.
    \item Inter-frame variance: The distortions at the same position of each frame are different, generating different noise patterns for each frame. 
    \item Intra-frame variance: Even in a single frame, the distortion patterns are spatially non-uniform.
\end{itemize}

One key observation is that the captured texts are subjected to occlusions (the missing or faded parts) caused by shot noise~\cite{shot} when there is an insufficient number of photons hitting the sensors. This can be easily reasoned in light of the short exposure time and small text pixel size causing reduced photons emitted and received. In addition,  other common imaging noise such as Gaussian noise gets visually amplified by relatively higher ISO values due to the bad light sensitivity of the webcam sensors. We call such noise ISO noise. Both two types of distortions have the potential to cause intra-frame and inter-frame variance.  The shot and ISO noise in the webcam peeking attack plays on a see-saw with an equilibrium point posed by the quality of the camera imaging sensors. It suggests that the threat level will further increase (see the comparison between the webcam and Nikon Z7's images in Figure \ref{fig:nnonuniform}) as future webcams get equipped with better-quality sensors at lower costs.

\subsection{Image Enhancing with MFSR.} \label{sec:mfsr}
The analysis of distortions calls for an image reconstruction scheme that can reduce multiple types of distortions and tolerate inter-frame and intra-frame variance. One possible method is to reconstruct a better-quality image from multiple low-quality frames. Such reconstruction problem is usually defined as multi-frame super resolution (MFSR)~\cite{yang2017image}. The basic idea is to combine non-redundant information in multiple frames to generate a better-quality frame. 

% The formulation of the MFSR problems often take into consideration various distortions to the images such as blur effects and other general noise.
% The reconstruction problem can be formulated as 

% \begin{equation}
% \widehat{I}_H = \min_I  \sum_{k=1}^{M}\left\|P_{k}\left(I\right)-I_{L}^{k}\right\|
% \end{equation}
% where $\widehat{I}_H$ is the estimated high-resolution image, $I^{k}_{L}$ is the $k$-th low-resolution image, and $P_k(\cdot)$ encodes various distortion terms. 
% % The comparison of performance limits between different MFSR approaches is challenging due to the lack of effective performance measure and benchmarks\cite{yang2017image,nelson2012performance}. We thus only utilize MFSR techniques as a practical method for improving the reflection image quality and avoid making definitive claims about the MFSR approaches. 

\begin{figure}[!t]
	\centering
	\includegraphics[width=.49\textwidth]{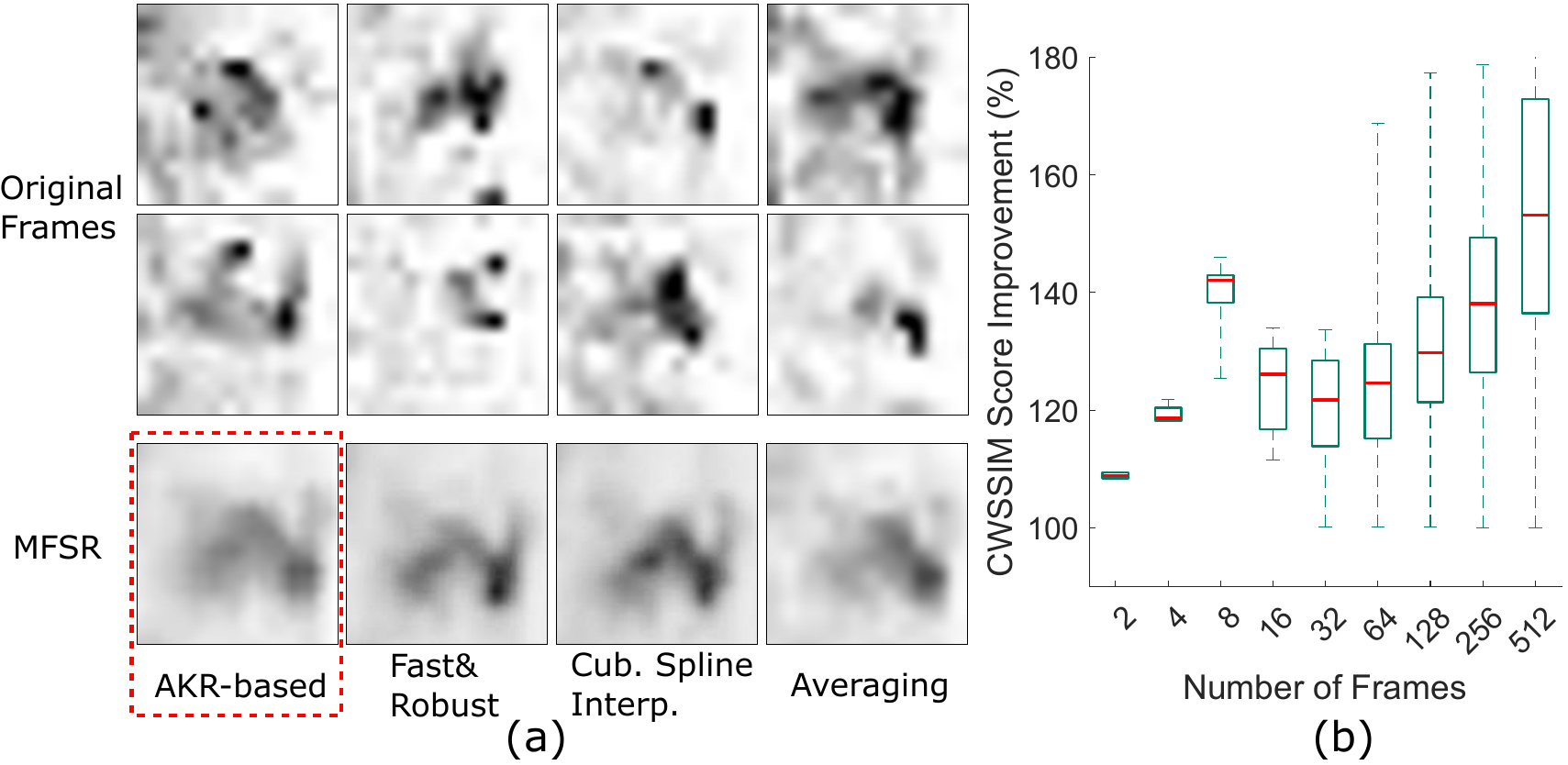}
	% \vspace{-.3in}
	\caption{(a) Comparison between single frames and the MFSR-reconstructed images with 4 different MFSR approaches. The MFSR images are reconstructed with the 8 frames shown at the top. The AKR-based approach generally produces the best reconstruction results in our task of reflection image reconstruction. (b) The improvement of reflection reconstruction quality as the number of frames used for MFSR increases.} 
	% \vspace{-.2in}
	\label{fig:W_ssr_comp}
\end{figure}

We tested 3 common light-weight MFSR approaches that do not require a training phase, including cubic spline interpolation~\cite{yang2017image}, fast and robust MFSR~\cite{farsiu2004fast}, and adaptive kernel regression (AKR) based MFSR~\cite{islam2009video}. Test results on the reflection images show that the AKR-based approach generally yields better results than the other two approaches in our specific application and setup. All three approaches outperform a simple averaging plus upsampling of the frames after frame registration, which may be viewed as a degraded form of MFSR. An example of the comparison between the different methods and the original 8 frames used for MFSR is shown in Figure \ref{fig:W_ssr_comp} (a). We thus use the AKR-based approach for the following discussions. 

One parameter to decide for the use of webcam peeking is the number of frames used to reconstruct the high-quality image. Figure \ref{fig:W_ssr_comp} (b) shows the CWSSIM score improvement of the reconstructed image over the original frames with different numbers of frames used for MFSR when a human wears the glasses to generate the reflections. Note that increasing the number of frames do not monotonically increase the image quality since live users' occasional intentional movements can degrade image registration effectiveness in the MFSR process and thus undermine the reconstruction quality. Based on the results, we empirically choose to use 8 frames for the following evaluations. In addition, the improvement in CWSSIM scores also validates that MFSR-resulted images have better quality than most of the original frames. We thus only consider evaluation using the MFSR images in the following sections.

\section{Reflection Recognizability \& Factors} \label{sec:eval}
In this section, we evaluate the recognizability limits of reflected texts enhanced by the MFSR method given a specific set of webcams, glasses, and advantageous environmental conditions. We then investigate the impact of the most significant  factors. The evaluations in this section are performed in a controlled lab environment and serve as the foundation for the  analysis  in Section \ref{sec:real}.

\begin{figure*}[!t]
	\centering
	\includegraphics[width=.88\textwidth]{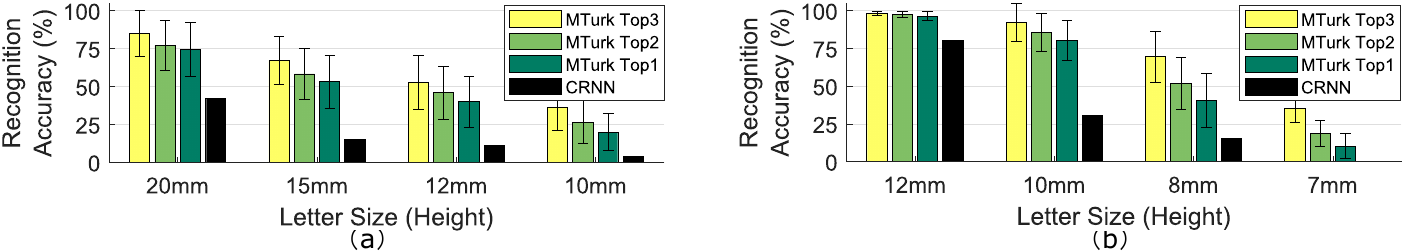}
	% \vspace{-.15in}
	\caption{The recognition accuracy of letters in different sizes with (a) the BLB glasses and (b) the prescription glasses. Although the pair of BLB glasses have higher reflectance than the prescription glasses, the prescription glasses enable reading smaller on-screen texts because of their smaller curvature leading to larger reflection pixel size. Note that the conclusion is device-specific and cannot be applied to general BLB-prescription glass comparison. Humans are found more capable of recognizing the reflected texts than SOTA OCR models.} 
	\label{fig:sizes_tops}
	% \vspace{-.2in}
\end{figure*}

\begin{figure*}[!t]
	\centering
	\includegraphics[width=.88\textwidth]{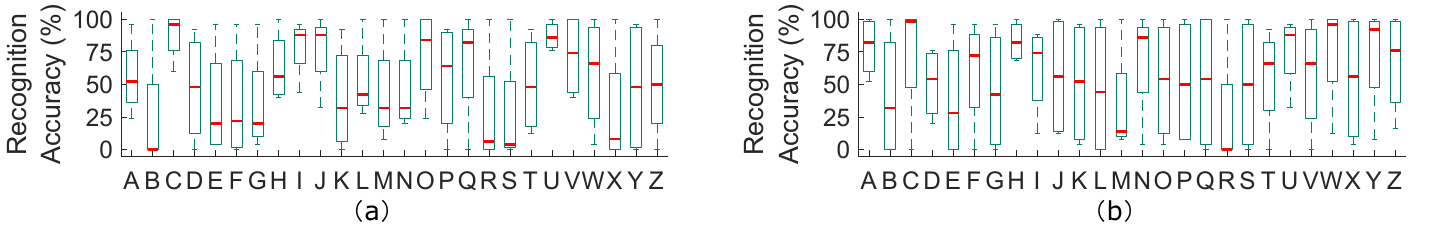}
	\vspace{-.1in}
	\caption{The human recognition accuracy of different letters with (a) the BLB glasses and (b) the prescription glasses. Letters such as ``R'' have been found the most difficult to read in the reflections while letters such as ``C'' and ``U'' have high recognizability. The difference is mostly due to  the simplicity and symmetry in the letters' structures which lead to smaller degradation of recognizability when the reflections are subject to distortions.} \vspace{-.1in}
	\label{fig:alphabet}
\end{figure*}

\subsection{Experimental Setup} \label{sec:exp_setup}

\textbf{Equipment.} We collected all data with the aforementioned Acer laptop as the victim device, and another Samsung laptop \cite{samsung} as the adversary's device. The two laptops were in a lab environment with WiFi network connection. The  victim laptop was measured to have an internet download speed of 246 Mbps and upload speed of 137 Mbps while those for the adversary laptop were 144 Mbps and 133 Mbps respectively. We used two pairs of glasses, i.e., the pair of BLB glasses and prescription glasses. 

\textbf{Data Collection.} We asked a person to wear the glasses and sit in front of the victim's laptop. The glass-screen distance was chosen to be 40 cm  which was also found to be close to the average distance in the user study (see Figure \ref{fig:user} (b)). The screen brightness was 100\%. To estimate the limits of recognition, we used an environmental light intensity of 100 lux to generate the best reflections. We then displayed single capital letters (26 letters) on the victim screen with different heights ranging from 20 mm to 7 mm. The victim and adversary laptops had a Zoom \cite{zoom} session with a video resolution of 1280$\times$720. For each display of the letters, we recorded a 3s video of the victim's images on the adversary's laptop. We then used 8 consecutive frames starting from 1s for MFSR reconstruction and generated one corresponding image for each video. We generated 208 images in total for the 2 glasses each with 4 different sizes.

\textbf{Recognizability Evaluation.} In order to evaluate the recognizability of the reconstructed single-letter images and avoid potential bias introduced by the authors' prior knowledge of the reflections, we acquired recognition accuracy by (1) using multiple SOTA pre-trained deep-learning OCR models including Google Tesseract and Keras CRNN, and (2) conducting a survey (Section \ref{sec:irb}) on Amazon Mechanical Turk (AMT) \cite{mturk}. For the AMT study, we collected answers from 25 crowdsourcing workers for each reconstructed image and thus collected 5200 answers in total. We showed to the workers all reconstructed images in a randomized manner without providing them with any information on the original letters on the screen. We asked the workers to provide 3 best guesses of the single letter in each reconstructed image. They were allowed to input the same answer for multiple guesses if they feel confident in a guess, or if they have no clue about making subsequent guesses.  
% We argue that this configuration helps avoid false increase in the recognition accuracy by making random guesses and thus show a lower bound of the recognition accuracy. The lower bound characteristic is also ensured by the fact that we provided no prior information to the workers. Theoretically, an experienced adversary may utilize her knowledge of the possible distortions to the images to increase the recognition success rate.
The recognizability of the texts in the reconstructed images is then represented by the recognition accuracy, i.e., correctly recognized number of letters over the total number of letters in each case.

% \begin{figure*}[!t]
% 	\centering
% 	\includegraphics[width=.98\textwidth]{figs/recognizability.pdf}
% 	\vspace{-.15in}
% 	\caption{\ly{Not sure if want to keep this fig.} Qualitative relationship between the reflection recognizability and the physical factors. The x-axes and y-axes of the sub-figures (a-h) represent the physical quantities and reflection recognizability respectively. The red dashed curves show the non-monotonic behaviors of webcams caused by auto-exposure control.} 
% 	\label{fig:recognizability}
% \end{figure*}

\subsection{Recognizability vs. Size \& Letter}

Figure \ref{fig:sizes_tops} shows the recognition accuracy with the BLB and prescription glasses respectively with different letter sizes. The AMT accuracy for each letter size is calculated by including all 25 answers for all 26 letters, i.e., with a denominator of $25\times26=650$. We picked 4 representative letter sizes for each pair of glasses respectively, and show the top 1, 2, and 3 recognition accuracy. we also use error bars to show the standard deviations. The SOTA OCR models performed considerably worse than AMT workers. \rev{We believe the main reason is that data distribution in the models' training sets is very different from the  actual  data in webcam peeking. After testing different image data on the models,  we found the two main causes for their bad performance are (1) significantly lower contrast, (2) occlusions caused by insufficient photons. Surprisingly, we also found  the models sensitive to how we crop the images, which suggests the convolutional layer features and potential data augmentation schemes employed by these models are not dealing well with our data’s distribution. We think future researchers can potentially utilize these pretrained models and collect their own webcam peeking dataset to fine-tune the model weights to better adapt machine learning recognition models to this scenario.}

The prescription glasses generally yield better results for the webcam peeking attack, showing that 10 mm texts can be recognized in the reconstructed images with over 75\% accuracy. Although not as good as the prescription glasses, the recognition accuracy with the BLB glasses is also high enough to support efficient peeking attacks against texts of 10-20 mm. Despite the better reflective characteristics of the BLB glasses, the prescription glasses still generate better results due to their smaller curvature, highlighting the risks of the peeking attack even without highly reflective glasses. 

Intuitively, different letters in the alphabet would be recognized with different levels of hardships due to their structural characteristics (see Figure \ref{fig:alphabet}).   For instance, the letters ``R'' and ``B'' have been found the hardest to recognize in both cases of the two pairs of glasses. On the other hand, letters such as ``C'', ``U'', ``I'', and ``O'' have generally the highest recognizability in all the sizes, which we suspect is due to their simple or highly symmetric structures that prevent the recognizability of such letters from dropping too seriously when the texts are down-sampled and occluded. Furthermore, we found letters having similar structures are confused with each other more easily in the recognition. For instance, ``J'' and ``L'' are mostly recognized as ``I'' when the letter size gets small because the distortions to the bottom part of ``J'' and ``L'' makes them just like ``I'' in the reflection images.

\begin{figure*}[!t]
	\centering
	\includegraphics[width=.99\textwidth]{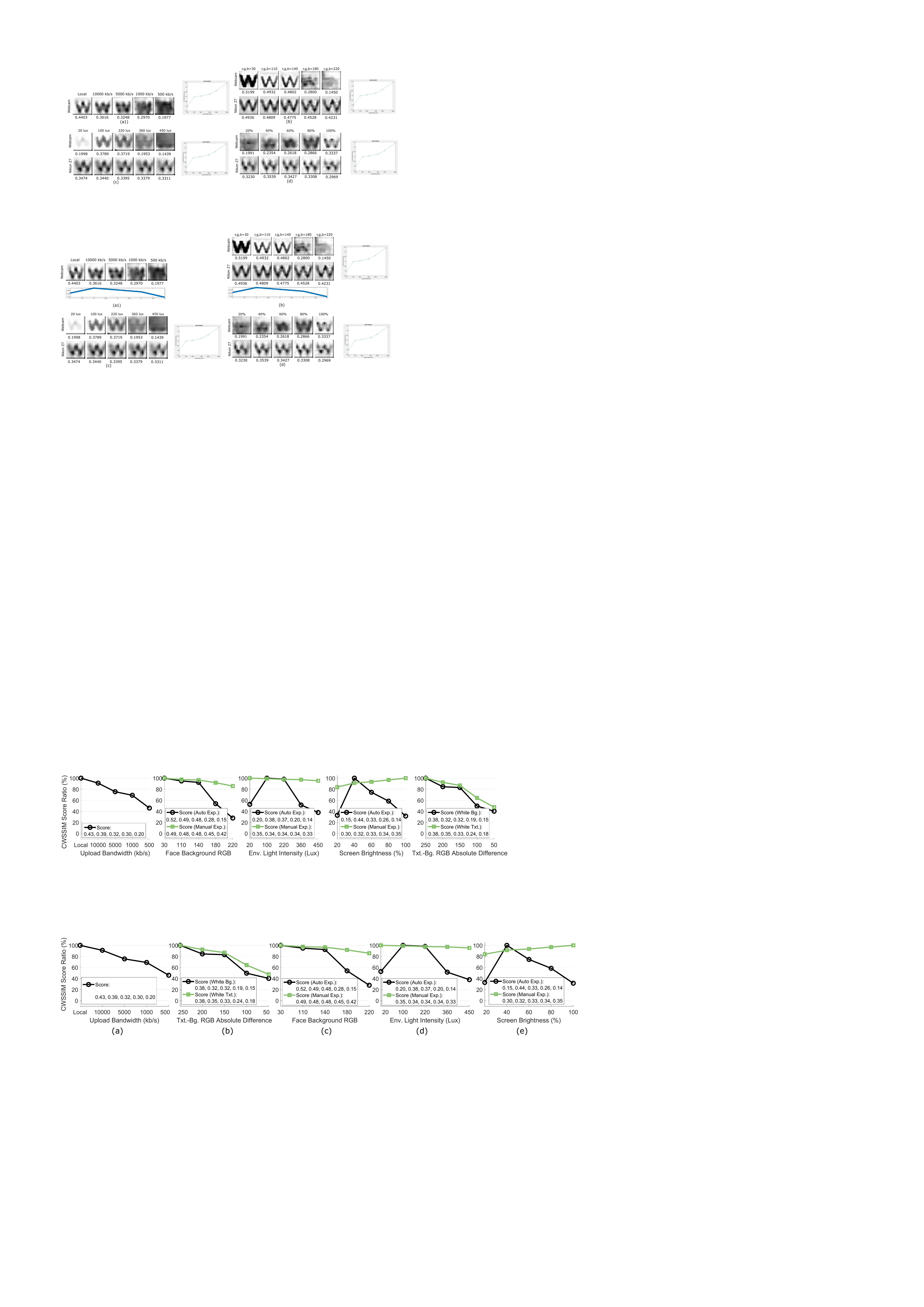}
	% \vspace{-.1in}
	\caption{Effects of impact factors evaluated by CWSSIM scores. The original score numbers are displayed along with the legend at the bottom, and we plot the ratio between each score and the highest score in each case as a percentage. Visualizations of the effects can be found in the appendix.}  
	\label{fig:factors_all}
	% \vspace{-.1in}
\end{figure*}

\subsection{Network Influence} \label{sec:zoom}

Video conferencing platforms like Zoom cause different levels of distortions in the images through video encoding and decoding under various network bandwidths. To analyze the impact, we compared the quality of the reconstructed images under different network bandwidths to that when the video is recorded by the victim's local device without going through Zoom. A visual demonstration of the effect is shown in Figure \ref{fig:factors_visual} which is quantified with CWSSIM scores and shown in Figure \ref{fig:factors_all} (a). We found that when the upload bandwidth is larger than 10 Mbps, the quality of the reconstructed images generally remains the same, and is close to the locally-captured and reconstructed images with a minor degree of added distortions. An upload bandwidth smaller than 10 Mbps starts to undermine the reconstructed image quality over Zoom. When the bandwidth is smaller than 1000 kpbs, the letters get hard to recognize. It's almost unrecognizable with a bandwidth smaller than 500  kbps. When the bandwidth was larger than 1500 kbps, Zoom was generally able to maintain a 720p video resolution with a frame rate close to 30 fps (Appendix \ref{apdx:zoom_bw}).

\subsection{Physical Factors} \label{sec:factors}
The recognizability of the reflections is a highly complex multi-variate function over many physical factors. We categorize the factors into 2 groups, namely those mainly affecting the reflection pixel size (Section \ref{sec:model}) and those affecting the light SNR. Comprehensive quantitative modeling of light SNR is very challenging due to the need for accurate imaging sensor models. Nevertheless, we provide qualitative analysis and quantify representative cases by calculating changes in CWSSIM scores (Figure \ref{fig:factors_all}).

In light SNR, the signal portion comes from the light emanating from the screen, reflected by the glasses, and then captured by the imaging sensors corresponding to the area of the screen. Other light captured by sensors in this area can be treated as noise. Counter-intuitively, more reflected light does not always lead to higher reflection recognizability as we will discuss next.
Figure \ref{fig:factors_all} (b-e) show the factors that can change light SNR most significantly. (c-e) also inspect how auto exposure  and manual (fixed) exposure can affect the light SNR-recognizability relationships in surprisingly different ways by using the laptop built-in webcam and the configurable Nikon Z7 respectively. 

\textbf{Text Color Contrast.} Different colors of texts can affect the reflection recognizability because the texts and screen background colors produce a certain contrast. We found that chroma has smaller effects than luma and  show how luma affects reflection quality in Figure \ref{fig:factors_all} (b) (visualization in Figure \ref{fig:factors_visual} (b)) by using the absolute difference in RGB values of gray-scale text and background colors to represent the contrast. As expected, lower contrast (smaller RGB difference) undermines the reflection recognizability. 

\textbf{Face Background Reflectance.}  
Face background reflectance is decided by sub-factors such as skin color. We tested different background reflectance by pasting the inner side of the glasses with papers of different gray-scale colors that have the same values for RGB. When the background has a higher reflectance (larger RGB values), more light from the environment as well as the screen will be reflected by it, increasing the noise portion of the light SNR and thus undermining the recognizability of the reflections as shown in Figure \ref{fig:factors_all} (c) (visualization in Figure \ref{fig:factors_visual} (c)).  

\textbf{Environment Light Intensity.}
A decrease in the environmental light intensity causes a smaller degree of noise and thus increases the light SNR. This increase, however, does not necessarily lead to better recognizability in the case of webcams which often have auto-exposure control to adjust the overall brightness of the videos they take. When the overall environment is too dark, the webcam's firmware automatically increases the exposure time trying to compensate for the dark environment. This increase in the exposure time  can cause an over-exposure for the reflected contents on the glasses which could have much higher light intensity than the environment, leading to smaller contrast and thus harder-to-read images. Such over-exposure is found in multiple participants' videos in the user study (Section \ref{sec:user}). On the other hand, the recognizability monotonically increases in the case of manual-exposure cameras such as the Nikon Z7 in manual mode. Figure \ref{fig:factors_all} (d) (visualization in Figure \ref{fig:factors_visual} (d)) shows the different behaviors of auto and manual exposure.

\textbf{Screen Brightness.} Screen brightness is the opposite of environmental light intensity in terms of its impact on the reflection recognizability. When the screen is brighter, the signal portion in the light SNR increases and can lead to more readable reflections for manual-exposure cameras. However, auto-exposure of most webcams can again negatively affect recognizability. Specifically, if the screen gets too bright compared to the environmental lighting condition, the webcams will often adjust their exposure time and ISO based on the dominant environmental light condition, and thus cause over-exposure to the screen reflections. Figure \ref{fig:factors_all} (e) (visualization in Figure \ref{fig:factors_visual} (e)) shows the effects.

\textbf{Summary.} The results show that variations in physical conditions can change the actual limits of the attack dramatically. The fact that reflection recognizability does not change monotonically with some factors like environmental light intensity and screen brightness further challenges the attack by making it more difficult to predict the possible outcomes in uncontrolled settings.

\color{revcol}

% \begin{table}
% \centering

% \caption{\rev{Correlation Scores Between Lens Parameters and \\ Reflection Quality} }
% \begin{tabular}{|c|c|} 
% \hline
% \textbf{Lens Parameter} & \begin{tabular}[c]{@{}c@{}}\textbf{Corr}\\\textbf{Score}\end{tabular} \\ 
% \hline
% Focal Length & 0.56 \\ 
% \hline
% Prescription Strength & 0.42 \\ 
% \hline
% Reflectance & 0.32 \\ 
% \hline
% Surface Coating Condition & 0.31 \\ 
% \hline
% Green/Red Spectrum Ratio & -0.08 \\ 
% \hline
% Blue/Red Spectrum Ratio & -0.02 \\ 
% \hline
% Blue/Green Spectrum Ratio & 0.14 \\
% \hline
% \end{tabular}

% \label{tab:lensparam}
% \end{table}

\subsection{Eyeglass Lens} \label{sec:fac_lens}
The difference in recognition accuracies between the pair of BLB and prescription glasses (Figure \ref{fig:sizes_tops}) suggests parameters of different eyeglass lenses will influence the performance of webcam peeking. To examine the impact, we analyzed 16 pairs of eyeglasses by inspecting the correlation between their reflection quality quantified by CWSSIM scores and several lens factors.  The CWSSIM scores are acquired with the 16 glasses when all other factors are kept the same.   

The results suggest lens focal length, which determines the pixel size of reflections (Equation~\ref{eq:convex}), has the strongest influence on the reflections with a correlation score of 0.56. The minimum, mean, and maximum focal length of the 16 pairs of glasses are 10, 268, and 110 cm respectively. With a correlation score of 0.42, the second strongest factor is found to be prescription strength (lens power) as lens power usually has a positive correlation with focal length following design conventions (see Appendix~\ref{apdx:dev} for explanation). Lens reflectance and surface coating conditions that mainly affect reflection light SNR produce correlation scores of 0.32 and 0.31 respectively. We empirically defined and added the factor of lens coating condition that gauges how much the lens coatings have worn off with higher values representing more intact coating. The motivation is our observation that damage in lens coating reduces the recognizability of reflections (see Figure~\ref{fig:glassdemo}). We also estimated lens reflection spectrum by calculating the ratio between RGB values of the reflections in the image but only found correlation scores lower than 0.15. This suggests the glass type (e.g., BLB or non-BLB) does not have a strong influence on reflection quality. Finally, we expect the parameters analyzed above have certain relationships with lens and coating materials, which require specialized optical equipment to measure and determine.

\color{black}

\section{\rev{Cyberspace Textual Target Susceptibility}} \label{sec:real}
The evaluations so far are based on the text's physical size and carried out in  controlled environments to better characterize user-independent components of the reflection model as well as the range of  theoretical limits for webcam peeking. \rev{In this section, we start by mapping the limits to common cyberspace objects in order to understand the potential susceptible targets. We then conduct a 20-participant user study with both local and Zoom  recordings to investigate the feasibility and challenges of peeking these targets and various factors' impact}.

\subsection{Mapping Theoretical Limits to Targets}
We use web texts as an enlightening example of cyberspace textual targets considering their wide use and the relatively mature conventions of HTML and CSS.  The discussion is based upon (1) a previous report \cite{scrap} scraping the most popular 1000 websites on Alex web ranking \cite{alexa}, and (2) a manual inspection of 117 big-font websites archived on SiteInspire \cite{bigweb}. We further divide the inspected web texts into 3 groups ($\mathcal{G}_1$, $\mathcal{G}_2$, $\mathcal{G}_3$, see Appendix~\ref{apdx:webtexts} and Table \ref{tab:webfonts}) in order to discuss separately how the webcam peeking attack with current and future cameras could have effects on them. As pointed out in Section \ref{sec:model}, the conversion between digital point size and physical cap height is dependent on specific user settings such as browser zoom ratio. The cap height values in Table \ref{tab:webfonts} are thus measured with the Acer laptop with default OS and browser settings as a case study.

\begin{figure*}[!t]
	\centering
	\includegraphics[width=.93\textwidth]{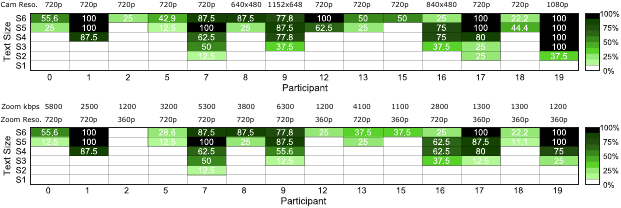}
	% \vspace{-.15in}
	\caption{The recognition results of textual reflections collected with local and Zoom-based remote video recordings from 20 user study participants. Participants 4, 14, and 3, 6, 10, 11  did not generate glass reflections that allow successful recognition due to problems of out-of-range viewing angles and very low light SNR respectively and are thus omitted from the figure.} 
	\label{fig:user_accs}
\end{figure*}
% \vspace{-.3in}

Based on the results in Figure \ref{fig:sizes_tops}, we hypothesize that the smallest cap heights adversaries can peek using mainstream 720p cameras is 7-10 mm. We then calculate the corresponding limits with 1080p and 4K cameras with Equation \ref{eq:convex} and show them in the Theoretical column of Table \ref{tab:webfonts}. Considering participants are most likely to use 720p cameras, we then choose point sizes S1-S6 in Table \ref{tab:webfonts} for evaluations.

\subsection{User Study} \label{sec:user}
The user study (Section \ref{sec:irb}) is designed in the following challenge-response way: An author generates HTML files each with one randomly selected headline sentence containing 7-9 words~\footnote{Uniform lengths (e.g., all 8 words) are avoided to prevent the adversary from guessing the words by knowing how long the sentences are.} from the widely-used ``A Million News Headlines'' dataset \cite{headline}. Only each word's first letter is capitalized. The participants display the HTML page in their browsers when they are recorded, and another author acting as the adversary tries to recognize the words from the videos containing the 20 participants' reflections without knowing the HTML contents by using the same techniques as in Section \ref{sec:eval}. We then calculate the percentage of correctly recognized words.

\begin{figure*}[!t]
	\centering
	\includegraphics[width=.9\textwidth]{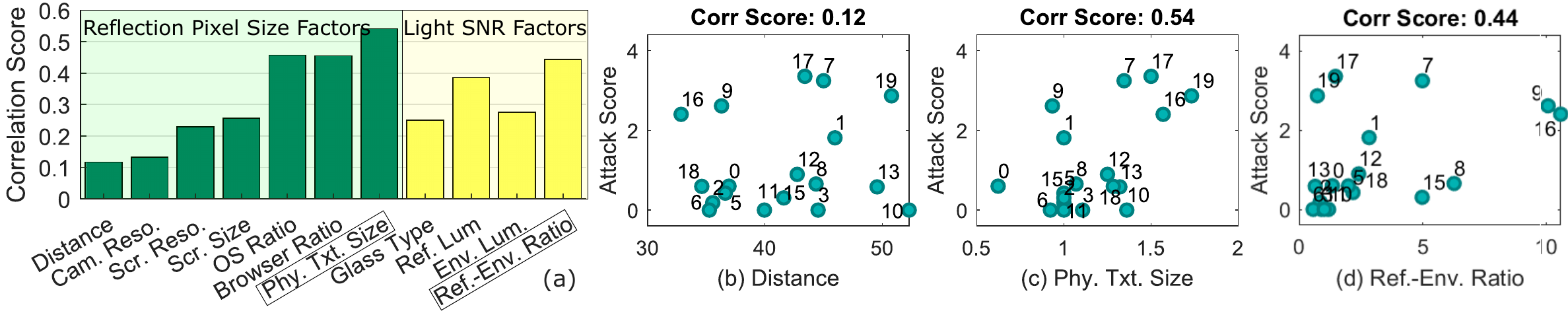}
	% \vspace{-.15in}
	\caption{(a) The degree of influence of different factors on the reflection recognition performance evaluated by the correlation scores. Factors highlighted with boxes are computed with other raw factors according to our model. (b-d) The joint distribution of three factors and the recognition results.} 
	\label{fig:user}
\end{figure*}

\textbf{Data Collection.} Each participant was given 6 HTML files of increasing point sizes from S1 to S6 as shown in Table \ref{tab:webfonts}. Note that the 6 sizes are specified in point size in HTML so that user-dependent factors such as screen size and browser zoom ratio can be studied (Equation \ref{eq:cap}). The participants display each HTML file on their own computer display in their accustomed rooms and behave normally as in video conferences. We allow participants to choose their preferred environmental lighting condition except asking them to avoid other close light sources besides the screen in front of their face. The reason is that we found a close frontal light source can seriously decrease light SNR, which can potentially be used as a physical mitigation against this attack but prevents us from examining the impact of all the other factors. \rev{We did not tell the participants to stay stationary and let them behave normally as in browsing screen contents.} Their webcams record their image for 30 seconds for each HTML.

\rev{Network bandwidth and resulted video quality are artifacts of video conferencing platforms that improve in a rapid way \cite{cisco} compared to other user-dependent physical factors. To study the present-day and possible future impact of video conferencing platforms, we record the 20 participants' videos both locally and remotely through Zoom. Our experiments focused on Zoom since it is the most used platform and also provides the most detailed video and network statistics.}

We asked the participants to report their user-dependent parameters including screen resolution ($N_{os}$), screen physical size ($H_{sr}$), OS and browser zoom ratio ($s_{os}, s_{b}$) webcam resolution in Equation \ref{eq:cap}, webcam resolution ($N$) in Equation \ref{eq:convex}, and the type of their glasses. Some other physical factors including environmental light intensity, screen brightness, glass-screen distance, and the physical size of displayed texts are difficult to be measured by the participants themselves and are not reported. We thus estimated the values of these factors by utilizing their videos.

\color{revcol}
\textbf{General Adversary Recognition Results.} The recognition results achieved by the adversary with local and remote recordings are shown in Figure \ref{fig:user_accs} (upper and lower respectively). Two participants (4 and 14) did not generate glass reflections of their screens in the video recordings due to the problem of out-of-range vertical viewing angles as predicted in Section \ref{sec:model}. Four participants (3, 6, 10, 11)  yield 0\% textual recognition accuracy due to a very low light SNR.

With local video recordings,  the percentage out of the 20 participants that are subjected to  non-zero recognition accuracy against S6-S1 are 70\%, 60\%, 30\%, 25\%, 15\%, and 0\% respectively. Videos of participants 7 and 17 using 720p cameras allowed the adversary to achieve 12.5\% and 25\% accuracies on recognizing S2. Videos of participant 16 using a 480p camera allowed the adversary to achieve an 37.5\% accuracy on recognizing S3.  These results translate to the predicted susceptible targets with cameras of different resolutions as listed in the User column of Table \ref{tab:webfonts}, where 720p webcams pose  threats to large-font webs ($\mathcal{G}_3$) and future 4K cameras pose threats to various header texts on popular websites ($\mathcal{G}_1$ and $\mathcal{G}_2$). As expected, this result is  worse than the theoretical limits in the table that are derived with prescription glass data in the controlled lab setting (Section~\ref{sec:eval}). Our observations suggest the main reasons include: (1) The environmental lighting conditions of the users are more diverse and less advantageous to screen peeking than the lab setup, generating reflections with worse light SNR. (2) Texts in the user study are mostly lower-case and have thus smaller physical sizes than the upper-case letters used in Section \ref{sec:eval}.  (3) The prescription glasses used in Section~\ref{sec:eval} have a larger focal length than the average user's glasses. (4) More intentional  movements exist in the user study leading to more motion blur. 

With Zoom-based remote recordings, the percentage of participants with non-zero recognition accuracy against S6-S1 degraded to 65\%, 55\%, 30\%, 25\%, 5\%, and 0\% respectively. We logged the video network bandwidth and resolution reported by Zoom as shown in Figure \ref{fig:user_accs}. The correlation between Zoom bandwidth, resolution, and their impact on video quality agrees with the observations in Section \ref{sec:zoom}. Generally, bandwidths smaller than 1500 kbps led to 360p resolutions for most of the time and  decreased the recognizable text size by 1 level. Zoom's 720p videos also caused degradation in recognition accuracy but mostly kept the recognizable text size to the same level as the local recordings, suggesting the same predictions of susceptible text sizes and corresponding cyberspace targets. 

Besides the mostly used platform Zoom, we also acquired remote recordings of participant 19 with  Skype and Google Meet. The adversary achieved better results with Skype than Zoom by recognizing S3 and S2 with 89\% and 25\% accuracies respectively, which is likely due to Skype's capability of maintaining better-quality video streams with a 1200 kbps bandwidth. The web-based Google Meet platform provided the lowest quality videos and only allowed the adversary to achieve 22\% accuracy on recognizing S4.

\color{black}

\textbf{Underlying Reasons.} To find out the dominant reasons enabling easier webcam peeking by analyzing the correlation between the recognition results and different factors, we turn each participant's results (6 sizes) into a single \textit{attack score} that is a rectified weighted sum of the recognition accuracy of the six text sizes tested. Figure \ref{fig:user} (a) shows correlation scores with 11 factors that affect reflection pixel size (left) and light SNR (right) respectively when $w=1.5$. The glass type includes prescription (15/20) and prescription with BLB coatings (5/20). The physical text size and reflection-environment light ratio highlighted in the boxes are two composite factors. In short, the physical text size represents the ratio between the actual physical size of texts displayed on each participant's screen and the case study values in Table \ref{tab:webfonts} and is calculated with Equation \ref{eq:cap} with other raw factors such as browser zoom ratios. The reflection-environment light ratio represents how strong the screen brightness is compared to the environmental light intensity and is calculated by dividing glass luminance by environmental luminance. Basically, these two composite factors represent our model's prediction of reflection pixel size and light SNR and are found to generate higher correlation scores than the other raw factors, which validates the effectiveness of our models. Figure \ref{fig:user} (b-d) further show the joint distribution of the attack score and three representative factors. It can be seen from (b) that the 40 mm screen-glass distance used in the evaluation of Section \ref{sec:eval} is about the average of the participants' values, and distances of these participants actually only have a very weak correlation with the easiness of webcam peeking attack. 
Figure \ref{fig:user} (d) suggests that when the screen brightness-environmental light intensity ratio gets lower than a certain threshold, the likelihood of preventing adversaries from peeking is very high, which may be considered as a temporary mitigation.

% Figure \ref{fig:user} (a-d) plot the joint distribution of the variables and recognizability score where each point represents one participant when $w=1.5$ with their Pearson correlation scores. Figure \ref{fig:user} (e) plots the correlation scores changing with the small-size emphasis factor. Participant 4 and 14 are dropped in all plots due to their viewing angle problem which is not relevant to the variables. (a-b) verify that higher light SNR caused by lower environmental light intensity and higher screen brightness generally improves the reflection recognizability. Somewhat surprisingly, there exist only very weak correlations between the glass-screen distance and recognizability in the user study. The displayed text physical size has been found to be the most significant variable. (e) suggests that when the adversary cares more about small-size texts ($w$ increases to 10), the displayed text physical size might even play a more important role. It is worth pointing out that the correlations are generally weak (<0.65). We believe the most likely reason is that we could not acquire the parameters of the participants' glasses, i.e., the glass curvature and reflectance due to the difficulty of large-scale measurement. As previously shown in Section \ref{sec:eval}, these glass parameters are indeed very important factors that can decide the reflection recognizability. In addition, the actual reflection pixel size can be calculated if glass curvature is known. 

\color{revcol}
\section{Website Recognition}

The results so far suggest it may still be challenging for present-day webcam peeking adversaries with mainstream 720p cameras to eavesdrop on common textual contents displayed on user's screens.  During our experimentation, we observed that recognizing graphical contents such as shapes and layouts on the screen is generally easier than reading texts. Although shapes and layouts contain more coarse-grained information compared to texts, a webcam peeking adversary may still pose non-trivial threats by correlating such graphical information with privacy-sensitive contexts. This work further explored to which degree can a webcam peeking adversary recognize on-screen websites by utilizing non-textual graphical information. 

\textbf{Data Collection.} 10 out of the 20 participants in the user study participated in the website recognition evaluation. Following a similar methodology as in \cite{juarez2014critical}, we used the Alexa top 100 websites as a closed-world dataset. We only investigate the recognition of the home page of each website in this work.  \cite{juarez2014critical} shows that other pages of a website can also lead to the recognition of the website. We believe the easiness of recognizing a website using different pages is worth exploring in future works. The experiment followed a similar procedure as the textual recognition experiment in Section~\ref{sec:real}. For each participant, one author generates a unique random sequence of 25 websites for the participant to browse (10 seconds for each website) while another author acts as the adversary that analyzes the video recordings. Both local and Zoom-based remote recordings were obtained and  recognized by the adversary. The adversary was given the whole recording and was asked to match each segment of the video to a specific website out of the 100 websites in the correct order. A random guess naive adversary is supposed to have a success rate of about 1\%.  Note that some participants changed their environment and ambient lighting compared to the previous textual recognition experiment since the two experiments were conducted five months apart.  

\begin{figure}[!t]
	\centering
	\includegraphics[width=.42\textwidth]{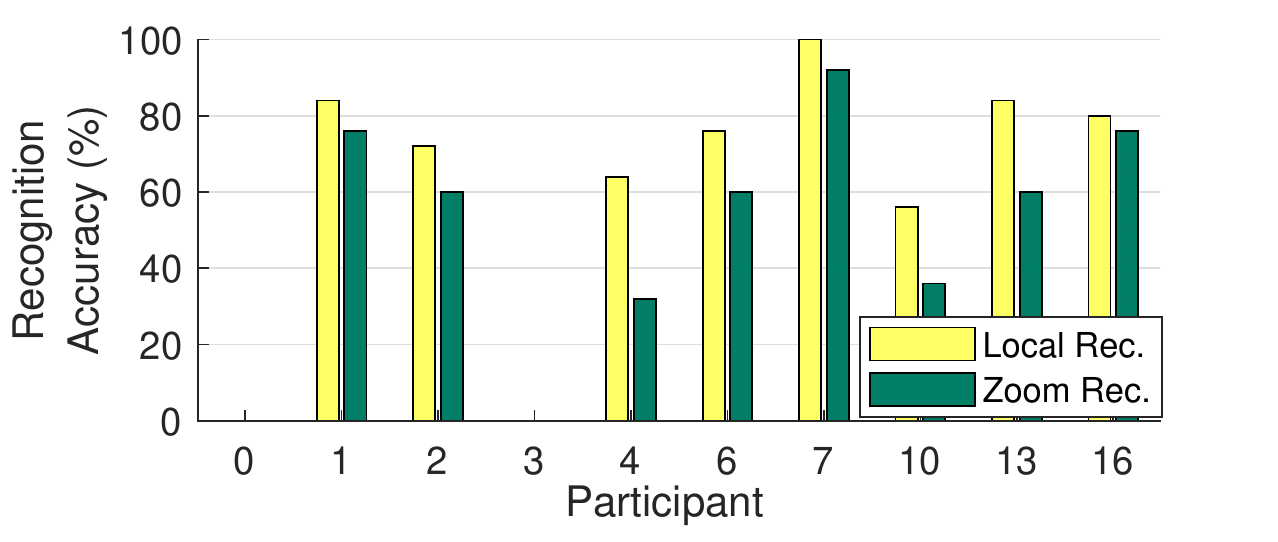}
	% \vspace{-.15in}
	\caption{\rev{Accuracy of recognizing Alexa top 100 websites from eyeglass reflections. Each participant browsed 25 websites.  Participant 0 and 4 did not yield recognizable reflections due to bad light SNR and viewing angles.}} 
	\label{fig:webrecog}
\end{figure}

\textbf{Recognition Results.} Figure \ref{fig:webrecog} shows the percentage of websites (out of 25) correctly recognized by the adversary. Participants 0 and 4 did not yield recognizable reflections due to bad light SNR and viewing angles respectively. This ratio of zero recognition (2  out of 10) agrees with that in the textual recognition test (6 out of 20), suggesting that webcam peeking may be impossible in 20-30\% video conferencing occasions due to extreme user environment configurations. 

As expected, participants with higher textual recognition accuracies such as participant 7 generally yield higher website recognition accuracies too. In addition, we observe that website recognition is more robust to various lighting conditions in the participants' ambient environment. For example, we found participant 10 who had 0\% textual recognition accuracy due to bad light SNR produced 56\% (local) and 36\% (remote) accuracies in website recognition with the same environment and lighting. The reasons are two-fold. First, solid graphical contents such as color blocks commonly found on web pages occupy larger areas than the body of texts and are thus much easier to identify in low-quality videos. Second, compared to black texts on white backgrounds which only have two different colors, the overall web pages with multiple graphical contents have more colors and contrast, leading to better robustness against over- and under-exposure of the usable screen contents in the webcam videos.

\textbf{Recognition Easiness and Web Characteristics.} Compared to texts, websites feature more abundant and diverse characteristics. We conducted qualitative and quantitative analyses to identify the characteristics that make certain websites more susceptible to webcam peeking. To that end, we ranked the 100 websites by their easiness of recognition utilizing recognition accuracies. Figure \ref{fig:webrank} shows rotated screenshots of the websites that rank the top and bottom 15 by their recognition easiness. Visual inspections suggest websites with higher contrast, larger color blocks, and more salient relative positions between different color blocks are easier to recognize. Websites that are mostly white with sparse textual and graphical components on them are the hardest to recognize. We calculated the correlation scores between the rank of each website and the average as well as the standard deviation of the websites' pixel values. Generally, a higher average means the website is closer to a pure white screen; a higher standard deviation means the website has more abundant high-contrast textures. The correlation scores obtained are -0.33 and 0.45.  

\color{black}

\section{Discussion}

\subsection{Proposed Near-Term Mitigations}
Given the threats, it is worthwhile exploring feasible mitigations that can be applied immediately. A straightforward approach involves users modifying the dominant physical factors identified in this work to reduce reflections'  light SNR, e.g., by placing  a lamp facing their face whose light increases the noise portion of light SNR. For software mitigations, we notice Zoom provides virtual filters of non-transparent cartoon glasses that can completely block the eye areas and thus eliminate reflections. Such features are not found in Skype or Google Meet. Other software-based approaches that support better usability involve fine-tuned blurring of the glass area. Although none of the platforms supports it now, we have implemented a real-time eyeglass blurring prototype that can inject a modified video stream into the video conferencing software. The prototype program\footnote{Details and open-source code of this prototype implementation can be found  at \href{https://github.com/longyan97/EyeglassFilter}{https://github.com/longyan97/EyeglassFilter}.} locates the eyeglass area and apply a Gaussian filter to blur the area. Figure~\ref{fig:filter} demonstrates the effect of using different strengths of Gaussian filtering by tuning the $\sigma$ parameter. Stronger filtering (higher $\sigma$) reduces reflection quality more but also undermines usability and user experience to a larger degree as it makes the users' eye areas look more unnatural. We believe the usable strength also depends on the characteristics of specific glasses. For example, Figure~\ref{fig:filter} shows three pairs of glasses with increasing reflectance. Since glasses with higher reflectance (e.g., the 3rd row) may already have produced screen reflections that occupy and distort images of users' eye areas, applying stronger filtering may cause less degradation in user experience in this case. On the other hand, lower-reflectance (e.g., the 1st row) glasses may require weaker filtering to maintain the same degree of usability. In general, we believe it is a good idea for future platforms incorporating this protection mechanism to allow users to adjust filtering strength by themselves.

\subsection{Improve Video-conferencing Infrastructure}

\textbf{Individual Reflection Assessment Procedure.} Our analysis and evaluation reveal that different individuals face varying degrees of potential information leakage when subjected to webcam peeking. Specifically, various factors of software settings, hardware devices, and environmental conditions affect the quality of reflections. Even for the same user, the potential level of threats varies when the user joins video conferences from different places or at different times of the day. These factors make it infeasible to recommend or implement a single set of protection settings (e.g., what glasses/cameras/filter strength to use) before the actual user settings are known. 

Providing usable security requires an understanding of how serious the problem is before trying to eliminate the problem. In light of this, we advocate an individual reflection assessment procedure that can potentially be provided by future video conferencing platforms. The testing procedure can be made optional to users after notifying them of the potential risk of webcam peeking. The procedure may  follow a similar methodology as the one used in this work by (1) displaying test patterns such as texts and graphics, (2) collecting webcam videos for a certain period of time, (3) comparing reflection quality in the video with test patterns to estimate the level of threats of webcam peeking. With the estimated level of threats, the platform can then  notify the user of the types of on-screen content that might be affected and offers options for protection such as filtering or entering the meeting with the PoLP principle that will be discussed below.

\textbf{Principle of Least Pixels.} Cameras are getting more capable than what average users can understand---unwittingly exposing information beyond what users intend to share.  The fundamental privacy design challenge with webcam technology is ``oversensing''~\cite{bolton2020curtail} where overly-capable sensors can provide too much information to downstream processing---more data than is needed to complete a function, such as a meaningful face-to-face conversation.  This oversensing leads to a violation of the sensor equivalent to the classic \textit{Principle of Least Privilege (PoLP)}~\cite{saltzer1975protection}. We believe long-term protection of users ought to follow a PoLP (perhaps a Principle of Least Pixels) as webcam hardware and computer vision algorithms continue to improve.   Thus, we recommend that future infrastructure and privacy-enhancing modules follow the PoLP not just for software, but for the camera data streams themselves. In sensitive conversations, the infrastructure could provide only the minimal amount of information needed and allow users to incrementally grant higher access privileges to the other parties. For example, PoLP blurring techniques might blur all objects in the video meeting at the beginning and then intelligently unblur what is absolutely necessary to hold natural conversations.

\subsection{User Opinion Survey}

We collected opinions on our findings of webcam peeking risks and expectations of protections from 60 people including the 20 people who participated in the user study and 40 people who did not. We did not find apparent differences between the two group's opinions.  The overall opinions are reported below.

\textbf{Textual Recognition.} For the discovered risk of textual recognition, 40\% of the interviewees found it a larger risk than what they expected; 48.3\% thought it was almost the same as their expectation; 11.7\% expected  worse consequences than what we found. In addition, 76.7\% of the interviewees think this problem needs to be addressed while 23.3\% think they can tolerate this level of privacy leakage.

\textbf{Website Recognition.} 61.7\% of the interviewees found it a larger risk than what they expected; 30\% thought it was almost the same as their expectation; 8.3\% expected  worse consequences than what we found. In addition, 86.7\% of the interviewees think this problem needs to be addressed while 13.3\% think they can tolerate this level of privacy leakage.

\textbf{Reflection Assessment.} Regarding the proposed idea of reflection assessment procedures that may be provided by video conferencing platforms in the future, 95\% of the interviewees said they would like to use it; 85\%, 68.3\%, 45\%, and 20\% of the 60 interviewees would like to use it when meeting with strangers, colleagues, classes, and family/friends respectively. 

\textbf{Glass-blur Filters.} Regarding the possible protection of using filters to blur the glass area, 83.3\% of the interviewees said they would like to use it; 78.3\%, 51.7\%, 43.3\%, and 11.7\%  of the 60 interviewees would like to use it when meeting with strangers, colleagues, classes, and family/friends respectively.

\subsection{Ethical Considerations}\label{sec:irb}

The AMT and user opinion survey received IRB waiver (No.HUM00208544) from the authors' institutes. The downloaded results are de-anonymized by only keeping their answers and deleting all other identifiable information including worker IDs. The results on the AMT and survey websites are deleted. We provided compensation of \$18/h for the workers. 

The textual and website recognition user studies are IRB-approved (No.ZDSYHS-2022-5). We ensured that participants and others who might have been affected by the experiments were treated ethically and with respect and anonymized participants with random orders. No personal information other than the videos and questionnaires was collected. The HTML files they used were created randomly by the authors and do not involve the participants' private information or contain any unethical or disrespectful information. The securely stored videos were used only for this research and not disclosed to third parties or used for other purposes.

\subsection{Limitation \& Future Work} 
 
This work used human-based recognition to evaluate the performance limits of reflection recognition. In future scenarios such as forensic investigations carried out by specialized institutions, we believe trained expert humans or machine learning methods may be employed to further increase the accuracy of reflection recognition. Compared to machine learning-based recognition, human-based recognition helps us understand the threats posed by a wide range of adversarial parties including even common users of video conferencing, and thus provides an estimate of the lower bound of the limits posed by camera hardware and other factors. We believe it is always possible to improve the attack performance by designing a more sophisticated machine learning model with more parameters,  increasing the size and diversity of the training dataset, etc.   Further, machine learning recognition is likely to face over-fitting and generalizability problems in webcam peaking due to highly varying personal environment conditions. Thus, we believe limits posed by a machine learning recognition back end are subjected to very large variances and require dedicated future works to quantify

Certain levels of biases have been introduced in the user study by informing the participants of the study's purpose. We envision that a future  study may conduct a real-world validation of this attack by performing it without participants' awareness while carefully following ethical regulations. Alternatively, public videos on social media may be analyzed to investigate how often such information leakage happens. A future study could also systematically interview professionals in different types of businesses and explore information leakage conditions, frequencies, and concerns. Contextual factors and user attitudes in real-world situations are complementary to this work's focus and worth investigating in future research.

\color{black}

\section{Related Work}\label{sec:teapot}

The problem of screen  reconstruction is a long-studied challenging problem. In this section, we analyze the past works that served as the foundations for our thinking in the context of video conferencing today and in the predicted future. 

\textbf{Screen Peeking Using Cameras.} Screen-peeking with cameras through optical emanation reflections has been explored in previous works. In 2008, Backes et al. \cite{backes2008compromising} showed that adversaries can use off-the-shelf telescopes and DSLR cameras to spy victims' LCD monitor screen contents from up to 30m away by utilizing the  reflective objects that can be commonly found next to the monitor screen such as teapots placed on a desk. In 2009, the authors \cite{backes2009tempest} took the attack to the next level by addressing the challenges of motion blur and out-of-focus blur by performing deconvolution on the photos with Point Spread Functions (PSF). Our work differs from these previous works by exploiting the victims' own webcams in video conferences for a remote attack. Such changes call for different imaging enhancing techniques due to the different types of image distortions. In addition, reflective objects on the desks and human eyes cannot be easily utilized due to very large curvatures. We thus exploit the glasses people wear to video conferences as a modern attack vector. \cite{weinberg2011still} proposed a relevant idea of using adversary-controlled webcams to detect changes in webpage links' colors for inferring visited websites. It requires the adversary to take control over the victim's webcam with malicious web modules and exploits coarse-grain color variations, while our work studies more natural attack vectors in video conferencing and investigate the limits of textual reconstruction.

\textbf{Screen Content Reconstruction With Other Emanations.} 
Besides the direct optical emanations from the screen that we exploit in this work, previous works also explored other channels such as electromagnetic radiation \cite{van1985electromagnetic,kuhn2004electromagnetic,kuhn2005security} and acoustic emanations \cite{genkin2019synesthesia}. Reconstructing screen contents with such emanations usually requires using additional eavesdropping hardware that is placed close to the victims by the adversary. On the other hand, our work exploits the victim's own webcams, making the attack more accessible.  

\textbf{Remote Eavesdropping Via Audio/Video Calls.}
Similar to our work, such attacks assume the adversary and victim are both participants of an audio/video conference, and the adversary can eavesdrop on privacy-sensitive information by analyzing the audio/video channels. For example, Voice-over-IP attacks for keystroke inference eavesdrop on the victim's keyboard inputs by utilizing timing and/or spectrum information embedded in the keystroke acoustic emanations \cite{cecconello2019skype,compagno2017don,shumailov2019hearing,elibol2012realistic}. Recently, Sabra et al. \cite{sabra2020zoom} proposed works solving the problem of inferring keystrokes by analyzing the dynamic body movements embedded in the videos during a video call. Hilgefort et al. \cite{hilgefort2021spying} spies victims' nearby objects through virtual backgrounds in video calls by carrying out foreground-background analysis and accumulating background pixels. In contrast, our work explores the related problem of  content reconstruction using only the optical reflections from participants' glasses embedded in the videos.

\section{Conclusion}
In this work, we characterized the threat model of the webcam peeking attack in video conferencing settings. We developed mathematical models that describe the relationship between the attack limits and different user-dependent factors. The analysis enables the prediction of future threats as webcam technology evolves. We conducted experiments both in controlled lab settings and with a user study. Results showed that present-day 720p cameras pose threats to the contents on users' screens when users browse certain big-font websites. Future 4K cameras are predicted to allow adversaries to reconstruct various header texts on popular websites. \rev{We also found adversaries can recognize the website users are browsing through webcam peeking with 720 cameras. We analyzed both short-term mitigations and long-term defenses and collected user opinions on the possible protections.}

\section*{Acknowledgement}

This work is supported by a gift from Analog Devices Inc. and China NSFC Grant 61925109 and 62201503. We thank our reviewers for their insightful comments that helped us improve this manuscript; we thank Dr. Cheng Yang for volunteering to wear eyeglasses in the lab experiments.

\bibliographystyle{plain}
\bibliography{refs}

\appendices

\color{revcol}

\section{Equipment Information} \label{apdx:dev}

\textbf{Lens Power \& Focal Length.} The power/Diopter of a lens is defined as the reciprocal of the lens' nominal  focal length. Different from the $f_g$ used before, this nominal focal length corresponds to the optical effect produced by the combination of the outer and inner surfaces of the lens, and is related to the radius of the outer and inner surfaces by the Lens Maker's Formula \cite{kuo2004sonic}:
$$
D = \frac{1}{f} = (n-1)(\frac{1}{R_o} - \frac{1}{R_i} )
$$
where $R_o$ and  $R_i$ are the radius of the outer inner surfaces respectively, and $n$ is the refractive index of lens material.  When the  lens power and materials are set, $R_o$ and  $R_i$ can both be adjusted to produce the desired power. However, flatter outer surfaces, as known as base curves, are often used for higher power lenses \cite{basecurve}. This is why we observe a positive correlation between $f_g$ and the lens power in Section~\ref{sec:fac_lens}. 

\color{black}

\textbf{Webcam Parameter Estimation.}  Manufacturers of the laptop built-in webcams often do not share information about the webcam focal length $f$ and imaging sensor physical size $W$. In this case, further estimation needs to be made. The term $\frac{f}{W}$ is a function of the vertical field-of-view (FoV) of the webcams. Specifically, the FoV angle $\alpha$ can be written as 
$$
\alpha=2 \tan ^{-1} \frac{W}{2 f}
$$
Considering that typical webcams have a diagonal FoV of in the range $70-90^\circ$, we can convert it to a typical vertical FoV of about $40-50^\circ$ for a 720p webcam and thus get $\frac{f}{W}$ approximately in the range of $1.1-1.4$ \cite{fov1,fov2,fov3}.

\textbf{Lab Setting Experiment Equipment.} The Acer laptop \cite{acer} has a screen width of 38 cm and height of 190 mm and a 720p built-in webcam. The OS is Ubuntu 20.04. The OS and browser zoom ratios are default (100\%). All the photos and videos are collected with the Cheese \cite{cheese}  webcam application. The photos are in PNG format and the videos are in WEBM format. The Samsung laptop used as the attacker device has OS Windows 10 Pro. The recordings are collected with OBS Studio in MP4 format.

The pair of BLB glasses \cite{blb_amazon} has lenses with a horizontal and vertical chord length of 5 cm and 4 cm respectively, and a focal length ($f_g$) of 8 cm. The pair of prescription glasses \cite{blb_amazon} has lenses with a horizontal and vertical chord length of 6 cm and 5 cm respectively, and a focal length of 50 cm.

Nikon Z7: The photos are in JPEG format (highest quality) and the videos are in MP4 format. We compared these formats with the compression-less (raw) photo and video formats provided by Nikon Z7 but didn't find an obvious difference in the image quality.

\begin{figure}[!t]
\centering
\includegraphics[width=.49\textwidth]{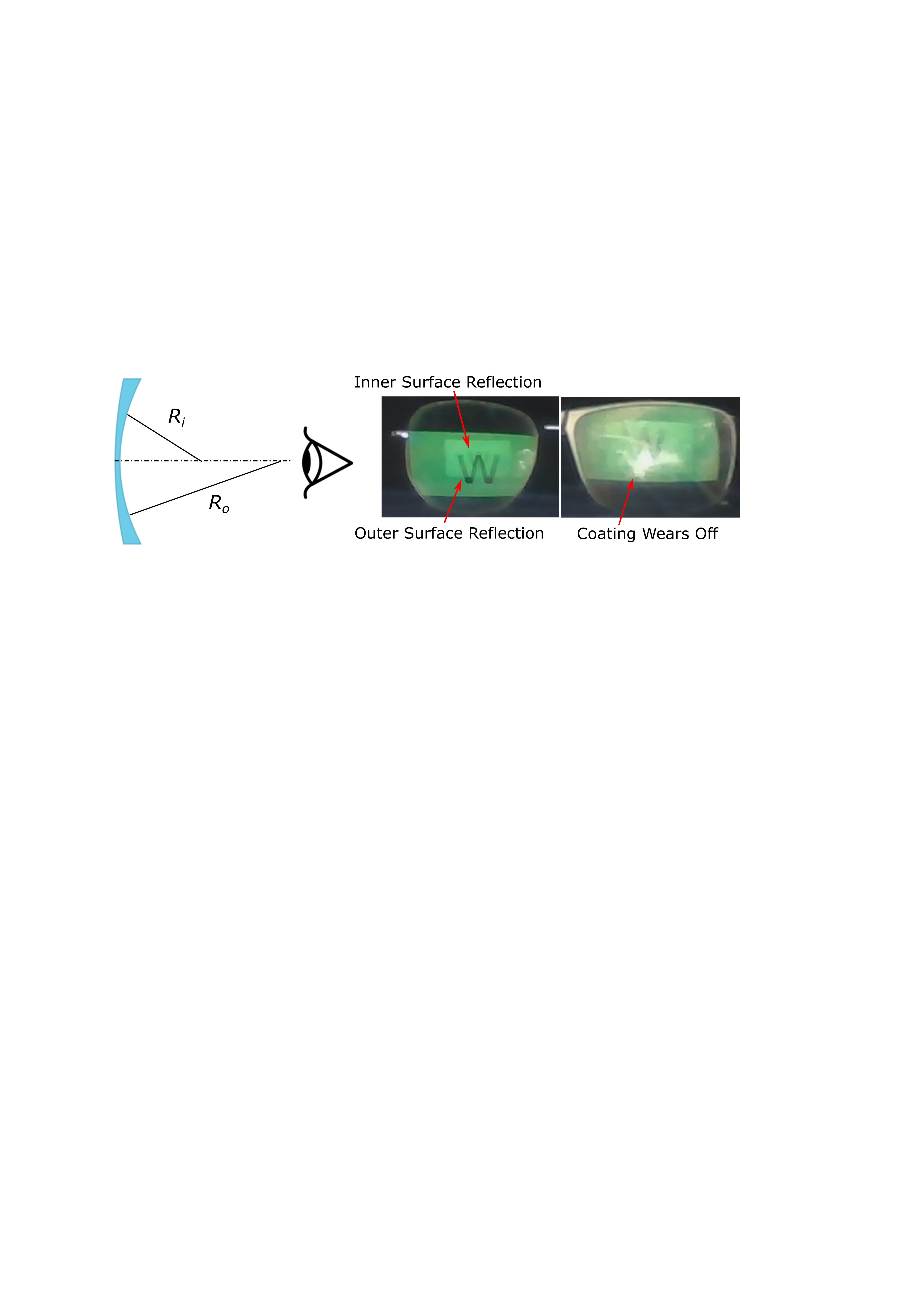}
\vspace{-.15in}
\caption{\rev{Design conventions suggest that eyeglasses with higher prescription strength have smaller curvature (larger radius/focal length) on the lens outer surface, leading to larger-size reflections. Besides curvature and reflectance, lens coating conditions can also affect reflection quality.}}
\label{fig:glassdemo}
\end{figure}

% \section{Objective Metrics for reflection quality} \label{apdx:metric}
% We embody this notion of reflection quality in the similarity between the reflected texts and the original templates. We compared multiple widely-used image structural and textural similarity indexes including structural similarity Index (SSIM) \cite{wang2004image}, complex-wavelet SSIM (CWSSIM) \cite{sampat2009complex}, feature similarity (FSIM) \cite{zhang2011fsim}, deep image structure and texture similarity (DISTS) \cite{ding2020image} as well as self-built indexes based on scale-invariant feature transform (SIFT) features \cite{lindeberg2012scale}. We found CWSSIM, which is robust to various rigid transforms, produces the best match with human perception results. We thus use the CWSSIM similarity score between reflections and templates as the reflection quality metric, which spans the interval $[0, 1]$ with larger numbers meaning more similar and thus higher quality.  

\section{Viewing Angle Model} \label{apdx:angle}

\begin{figure}[!t]
	\centering
	\includegraphics[width=.4\textwidth]{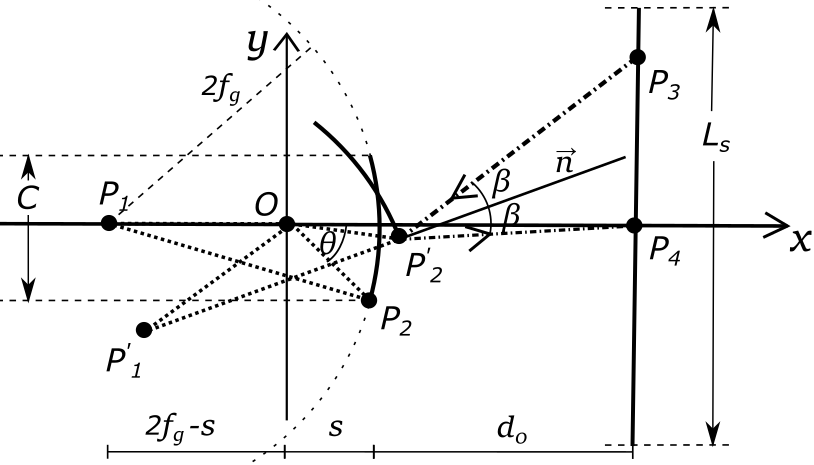}
    % \vspace{-.1in}
	\caption{The model of viewing angle.} 
	\label{fig:angle}
\end{figure}

Similar to the pixel size model, we only use 2D modeling (Figure~\ref{fig:angle}) for simplicity which can represent either horizontal or vertical rotations, and we only consider one glass lens since the two lenses are symmetric. The lenses are further modeled as spherical with a radius $2f_g$. We set the origin $O$ to the center of the head which is also treated as the rotation center, and assume the initial orientation without rotation is such that the center of the glass lens arc $P_1$ aligns with the rotation center and the laptop webcam $P_4$ on the X-axis. The distance between the glass lens center and the rotation center is $s$. To calculate the maximum feasible angles, we only need to consider the reflections from either one of the two boundary points of the glass lens since they are symmetric. We label the bottom boundary point as $P_2$. After a rotation of angle $\theta$, $P_1, P_2$ are rotated to $P'_1, P'_2$ respectively, and the vector $\overrightarrow{P'_1 P'_2}$ yields the normal $\vec{n}$ at the reflection point $P'_2$. $P_3$ denotes the point source on the screen whose light gets reflected to the camera with an incident angle $\beta$. With $L_s$ being the length of the screen on the dimension, the camera should be able to peek reflections from the glass lens if $P_3$ falls in the range of the screen. $C$ denotes the length of the glass lens chord.

In order to find a mapping from the rotation angle $\theta$ to the light-emission point $P_3$ on the screen, the key is to find the slope of the line $P'_2 P_3$ which intersects with the screen. Since $P'_1 P'_2$ bisects $P'_2 P_4$ and  $P'_2 P_3$ , we denote the slope of these three lines as $b_1, b_2, b_3$ respectively, and have

$$b_3 = \frac{b_2 - 2b_1-b_1^2b_2}{b_1^2 - 2b_1b_2 -1} $$

To calculate $b_1$ and $b_2$, the coordinate of $P'_1$ and $P'_2$, $P_4$ can be denoted as, 

$$\left\{\begin{array}{l} P'_1: \left((s-2f_g) cos\theta, (s-2f_g) sin\theta \right)  \triangleq (C,D)
\\ 
P'_2:\left(  x_0 cos\theta - y_0 sin\theta, x_0 sin\theta + y_0 cos\theta  \right)   \triangleq (A,B) 
\\
P'_2: \left( s+d, 0  \right)   \triangleq (E,0) 
\end{array}\right.$$
and thus 
$$ b_1 = \frac{B-D}{A-C}, \quad b_2 = \frac{B}{A-E}$$

The last missing piece is the coordinate of $P_2$, which is denoted as $P_2: (x_0, y_0) = (r\times cos\alpha, r\times sin\alpha)$, where 
$$\left\{\begin{array}{l} r = \sqrt{(\frac{C}{2})^2+(\sqrt{R^2-(\frac{C}{2})^2}-(R-s))^2}
\\
\alpha = -arcsin(\frac{C}{2r})
\end{array}\right.$$

We note that the measured ranges in Table~\ref{tab:angle} are uniformly larger than the theoretical values, which could be caused by a coarse estimation of the distance $s$ since the actual distance between the lens and the rotation center is hard to determine, and the fact that the model approximates the camera as a point instead of a surface.

\section{Video Conferencing Platform Behaviors} \label{apdx:zoom_bw}

% In Ubuntu, limit upload bandwidth using 
% sudo ./wondershaper -a wlp6s0  -u 2500

% check network bandwidth using iftraf -> detailed interface stats

\begin{figure}[!t]
	\centering
	\includegraphics[width=.35\textwidth]{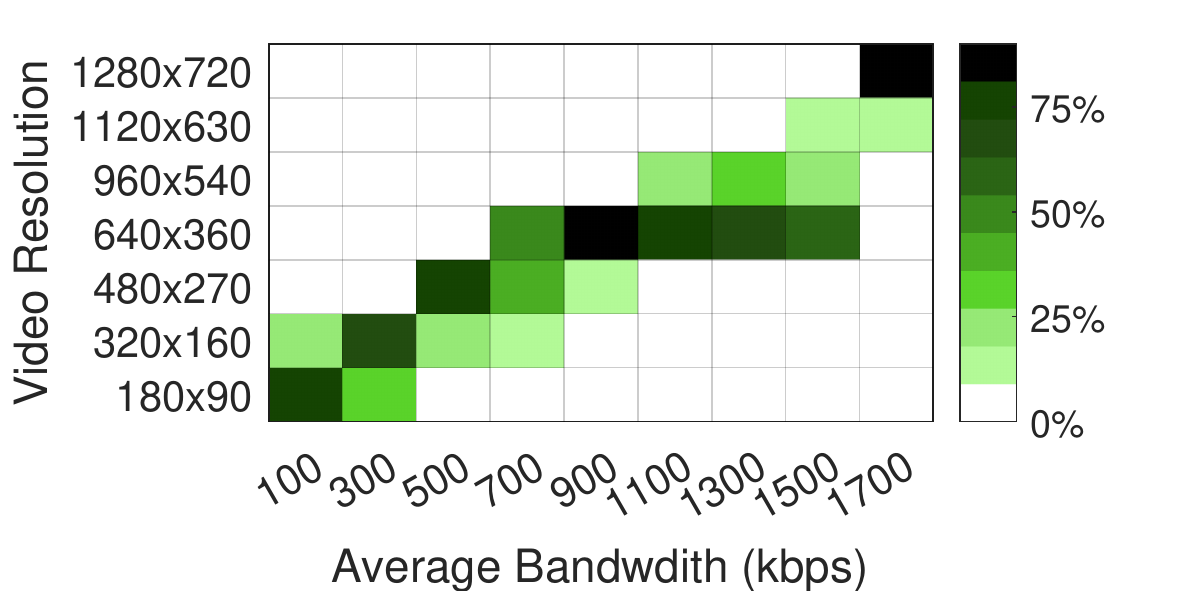}
	% \vspace{-.15in}
	\caption{\rev{Heat map of observed Zoom video resolutions under different low bandwidths that resulted in resolutions lower than 720p}} 
	\label{fig:zoombwreso}
\end{figure}

\textbf{Zoom Under Low Bandwidths.} When network bandwidth got smaller than 4 Mbps, we found Zoom will first experience a short period of aggravated packet loss, and then rapidly decrease the video resolution to compensate for it. Video resolution will soon be increased again by sacrificing frame rate as well as  compression loss. Zoom will still try to recover high frame rates later by further increasing the video compression loss. Through our experiments, we noticed that when the bandwidth was larger than 1500 kbps, Zoom was able to maintain a 1280*720 resolution with a frame rate very close to 30 fps. \color{revcol}  We observed lower resolutions when the bandwidth is lower than 1500 kbps, as shown in Figure~\ref{fig:zoombwreso}. Skype and Google Meet do not provide statistics like resolution, frame rate, and bandwidth. But our visual inspection suggests they take a similar approach as Zoom to handle bandwidth issues. 

\textbf{Video Quality Control.} Currently, Zoom and Skype do not provide an option for users to control video resolution or quality directly. Google Meet only allows users to switch between 720p and 360p send and receive video resolutions. However, users can limit their system or process bandwidths using software like NetLimiter to decrease video quality even without the conferencing platform offering such an option.

\color{black}

% \section{Visualization of Factors' Impact } \label{apdx:factors}
% Figure \ref{fig:factors_visual} shows the original reflections of Figure \ref{fig:factors_all}.

% \begin{figure}[!h]
% 	\centering
% 	\includegraphics[width=.4\textwidth]{figs/factors_visual.pdf}
% 	\caption{(a) The comparison between reconstructed images when the video is recorded locally on the victim device and over Zoom with different network upload bandwidths. (b) Changes of reflection recognizability with different text-background color contrast. (c) Changes of reflection recognizability with different background colors (reflectance). We tested gray-scale colors with the same RGB values, which have relatively uniform reflectance on the visible light spectrum. (d) Changes of reflection recognizability under different environmental light intensity. (e) Changes of reflection recognizability with different screen brightness.} 
% 	\label{fig:factors_visual}
% \end{figure}

\section{Distortion Analysis} \label{apdx:distortion}

\begin{figure}[!t]
	\centering
	\includegraphics[width=.48\textwidth]{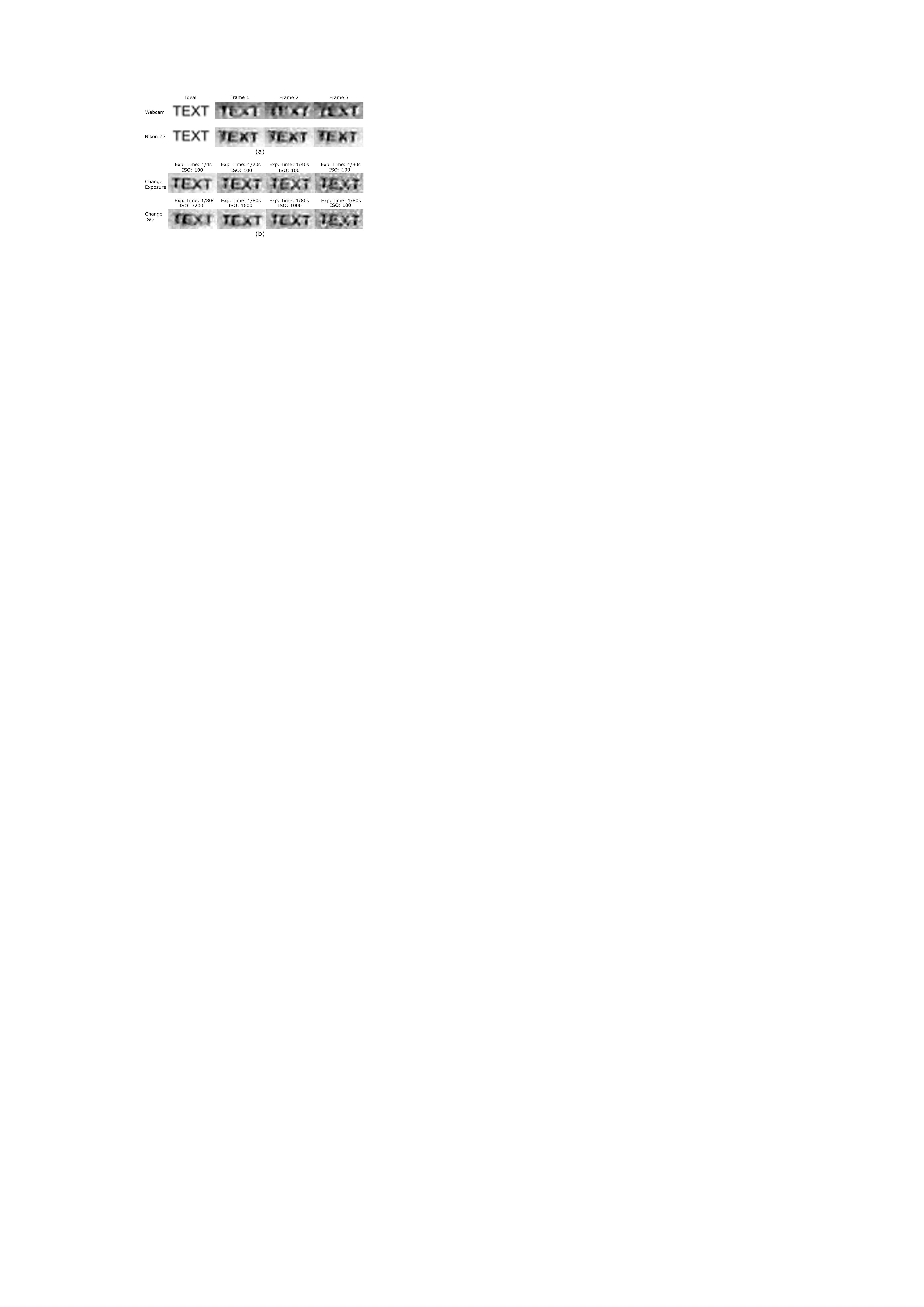}
	\vspace{-.1in}
	\caption{(a) The ideal capture versus the actual captures in three consecutive frames by webcam (1st row) and Nikon Z7 (2nd row). The distortions feature occlusions with inter-frame and intra-frame variance. The webcam yields larger variances. (b) Photos captured by Nikon Z7 under different exposure times and ISO settings. Longer exposure time and medium ISO yield smaller distortions and increase SNR.} 
	% \vspace{-.2in}
	\label{fig:nnonuniform}
\end{figure}

We taped the inner surface of the glasses lens with black papers in order to eliminate the impact of the face background and better characterize the inherent distortions. Effects of the face background are discussed in Section \ref{sec:factors}. The webcam and Nikon Z7 were set to the same color temperature (3500 K) and frame rate (30 fps). For the highly configurable Nikon Z7, we set the ISO, aperture, and exposure time to $100$, $F4$, and $\frac{1}{30}s$ respectively, disabled all active noise-reduction schemes including vibration reduction, and used manual focus mode. For both cases, we displayed the string ``TEXT'' and adjusted the size to make sure the captured text in both cameras' frames has a size of 10 pixels vertically.

Different from previous works~\cite{backes2008compromising,backes2009tempest}, motion blur and out-of-focus blurs that are theoretically uniform within a single frame is not the number one limiting factors in the webcam peeking threat model because of the relatively shorter exposure time and  closer, more  constant camera-object distance. Instead, distortions with intra-frame and inter-frame variance dominate which suggests the image quality cannot be easily improved with PSF deconvolution as in~\cite{backes2009tempest} and new image enhancing techniques are needed.

Figure \ref{fig:nnonuniform} (b) taken with the configurable Nikon Z7 shows how these two forms of distortions (shot and ISO noise) affect the images. For the first set of images (1st row), we keep ISO at $100$ and decrease the exposure time from $\frac{1}{4}s$ to $\frac{1}{80}s$ to show the effect of fewer photons hitting the image sensors which results in increased shot noise occlusions.  For the second set of images (2nd row), we keep the exposure time at $\frac{1}{80}s$ while increasing ISO from $100$ to $3200$ to show the effect of increased ISO noise.

\section{Web Textual Targets} \label{apdx:webtexts}

\begin{table}[!t]
    \centering
    
    \caption{Text sizes of web contents}
    % \vspace{-.1in}
    \begin{tabular}{|l|c|c|} \hline
    \textbf{Target} & \textbf{Point Size}  & \textbf{Cap Height (mm)}  \\ \hline
    $\mathcal{G}_1$ P & 12 &   2.1 \\ \hline
    $\mathcal{G}_1$ H3 & 14 &    2.5 \\ \hline
    $\mathcal{G}_1$ H2 & 18 &   3.2  \\\hline
    $\mathcal{G}_1$ H1 & 24 &    4.3  \\\hline
    $\mathcal{G}_2$ P & 21 &   3.7  \\\hline
    $\mathcal{G}_2$ H3 & 25 &   4.3  \\\hline
    $\mathcal{G}_2$ H2 & 32 &    5.6  \\\hline
    $\mathcal{G}_2$ H1 (S1) & 42 &   7.4  \\\hline
    $\mathcal{G}_3$ 0\% (S2)& 56 &   10  \\\hline
    $\mathcal{G}_3$ 20\% (S3)& 80 &   14  \\\hline
    $\mathcal{G}_3$ 40\% (S4)& 102 &  18 \\ \hline
    $\mathcal{G}_3$ 60\% & 136 &  24  \\\hline
    $\mathcal{G}_3$ 80\% (S5)&  253 & 35  \\ \hline
    $\mathcal{G}_3$ 95\% (S6)& 340  & 60  \\ \hline 

\end{tabular}
    \label{tab:webfonts}
\end{table}

\textbf{Web Text Design Conventions.} Despite the fact that the default CSS font sizes are decided by web browser vendors separately, we find many of them follow the W3C recommendation \cite{html4}, where H1, H2, H3 headers' font sizes are 2, 1.5, 1.17 em respectively. To briefly explain, a text size of $x$ em means the size is $x$ times the current body font size of the web page \cite{em} which is usually the same as the font size of paragraph (P) elements. Nevertheless, we note that web design standards are lacking and designers have a large degree of freedom of choosing their own text designs. Sometimes bigger fonts are preferred in order to make the websites more stylish and eye-catching. In this section, we thus investigate both conventional and more stylish web text sizes. 

\textbf{Text Sizes.} We summarize the  text sizes investigated in Table~\ref{tab:webfonts} where The cap height values are  measured with the Acer laptop and default OS and browser
settings.

$\mathcal{G}_1$ and $\mathcal{G}_2$: The first group represents the median HTML P, H1, H2, H3 texts of the 1000 websites. \cite{scrap} reports that the median size of the P elements is about 12 pt and H1, H2, H3 sizes are close to the 2, 1.5, 1.17 em ratios recommended \cite{html4}. We thus use these point sizes for $\mathcal{G}_1$ and specify the corresponding cap heights in Table \ref{tab:webfonts}. The second group represents the largest HTML P, H1, H2, H3 texts of the 1000 websites in \cite{scrap} with the same recommended em ratios for the headers. \cite{scrap} finds that about 4\% of the 1000 websites use a P size as large as 21 pt. This results in H1, H2, H3 sizes of 25, 32, and 45 pt respectively.

$\mathcal{G}_3$: The third group represents the 117 big-font websites' texts. We manually inspected all the 427 websites archived on SiteInspire\cite{bigweb}. The reason for manual analysis rather than scraping is that many large-font texts on the websites are embedded in the form of images instead of HTML text elements in order to create more flexible font styles. We then selected 117 of them based on the following criteria: (1) The webpage is still active. (2) The largest static texts that enable an adversary to identify the website through google search have a cap height of at least 10 mm when displayed on the Acer laptop. We show the different quantiles of the largest physical cap heights on the 117 websites and the converted point sizes in Table \ref{tab:webfonts}. We find that most websites in $\mathcal{G}_3$ are related to art, design, and cinema industry which like to present their stylish design skills but unfortunately make the web peeking attack easier. About 1/3 of the websites are designers' or studios' websites that computer science/security researchers may overlook.  Furthermore, 72 out of the 117 websites are ranked on Alexa from 38 to 8,851,402 with 5 websites among the top 10,000.

% \begin{figure*}[!t]
% 	\centering
% 	\includegraphics[width=.88\textwidth]{figs/alphabet.pdf}
% 	\vspace{-.1in}
% 	\caption{The human recognition accuracy of different letters with (a) the BLB glasses and (b) the prescription glasses. Letters such as ``R'' have been found the most difficult to read in the reflections while letters such as ``C'' and ``U'' have high recognizability. The difference is mostly due to  the simplicity and symmetry in the letters' structures which lead to smaller degradation of recognizability when the reflections are subject to distortions.} \vspace{-.1in}
% 	\label{fig:alphabet}
% \end{figure*}

\begin{figure}[!t]
	\centering
	\includegraphics[width=.40\textwidth]{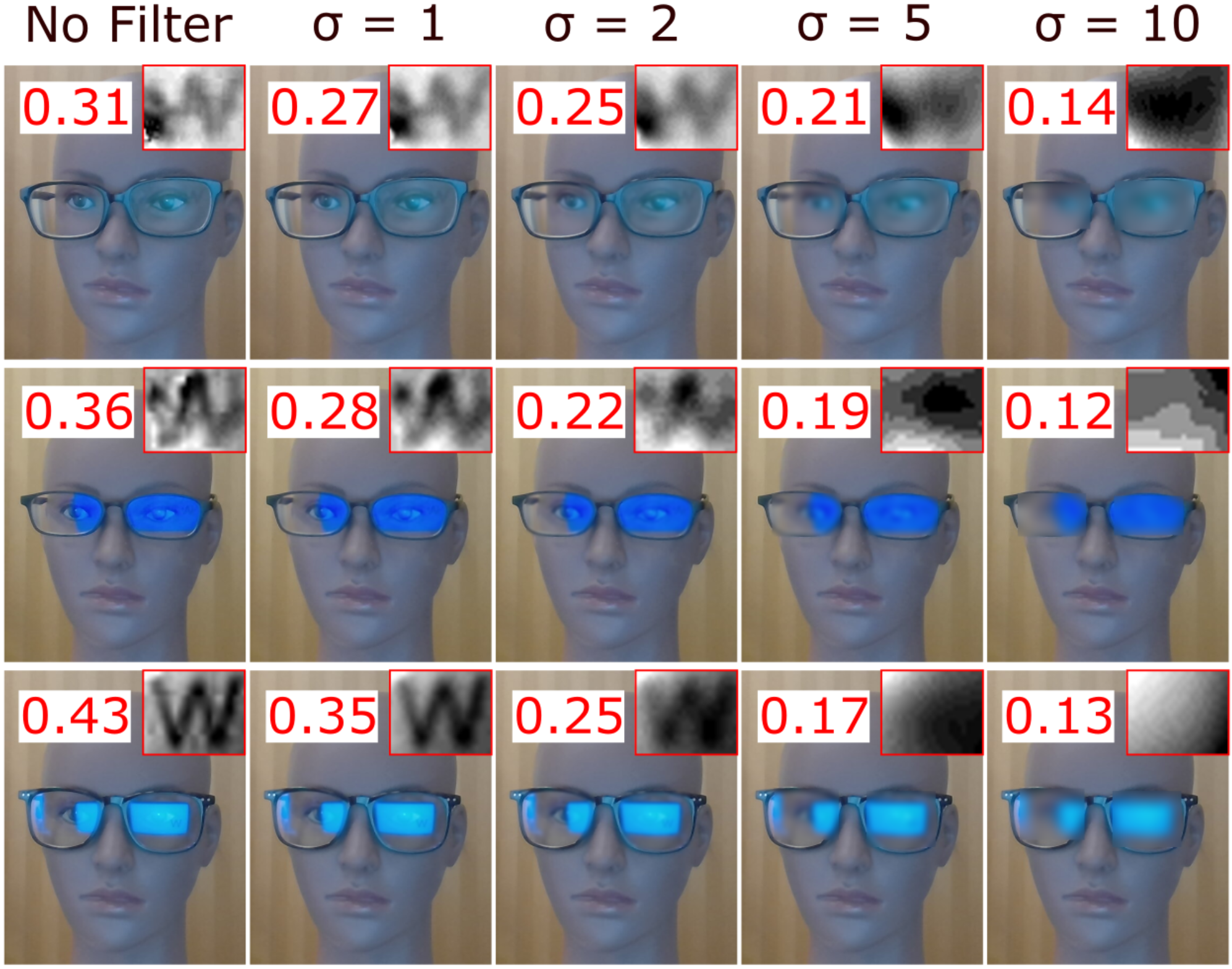}
	\vspace{-.1in}
	\caption{Different strengths of Gaussian filtering  applied on three pairs of glasses. The reflected texts and their CWSSIM scores in each case are shown. Different glasses require different strengths of filters to reduce the reflection. We thus advocate an individual reflection testing procedure to determine protection scheme and settings.}
	\label{fig:filter}
\end{figure}

% \section{Human Subjects Research Information} \label{apdx:irb}
% \textbf{AMT.} The AMT study received IRB waiver from the authors' institutes. The survey results downloaded from AMT website are de-anonymized by only keeping their answers and deleting all other information including worker IDs. The results on the AMT website are deleted. We provided compensations of \$18/h for the workers. 

% \textbf{User Study.} The participants were anonymized with random orders. No personal information other than the videos and questionnaires was collected. The HTML files they used were created randomly by the authors and do not involve the participants' private information or contain any unethical or disrespectful information. The participants' videos were used only for this research and not disclosed to third parties or used for other purposes.

\begin{figure}[!t]
	\centering
	\includegraphics[width=.45\textwidth]{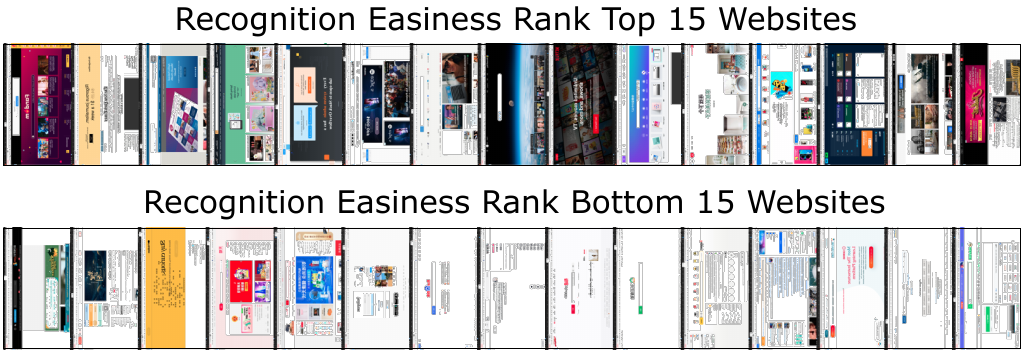}
	\vspace{-.1in}
	\caption{\rev{A spectrum of Alexa top 100 websites that are found to be the easiest (upper) and hardest (lower) to recognize in our evaluation of website recognition under webcam peeking attacks. Screenshots of each website are rotated by 90 degrees and concatenated horizontally. Correlations scores between the rank of website recognition easiness and website pixel values' average and standard deviation are -0.33 and 0.45 respectively, suggesting darker websites with high-contrast graphical contents are easier to recognize.}} 
	\label{fig:webrank}
\end{figure}

\begin{figure}[!t]
	\centering
	\includegraphics[width=.4\textwidth]{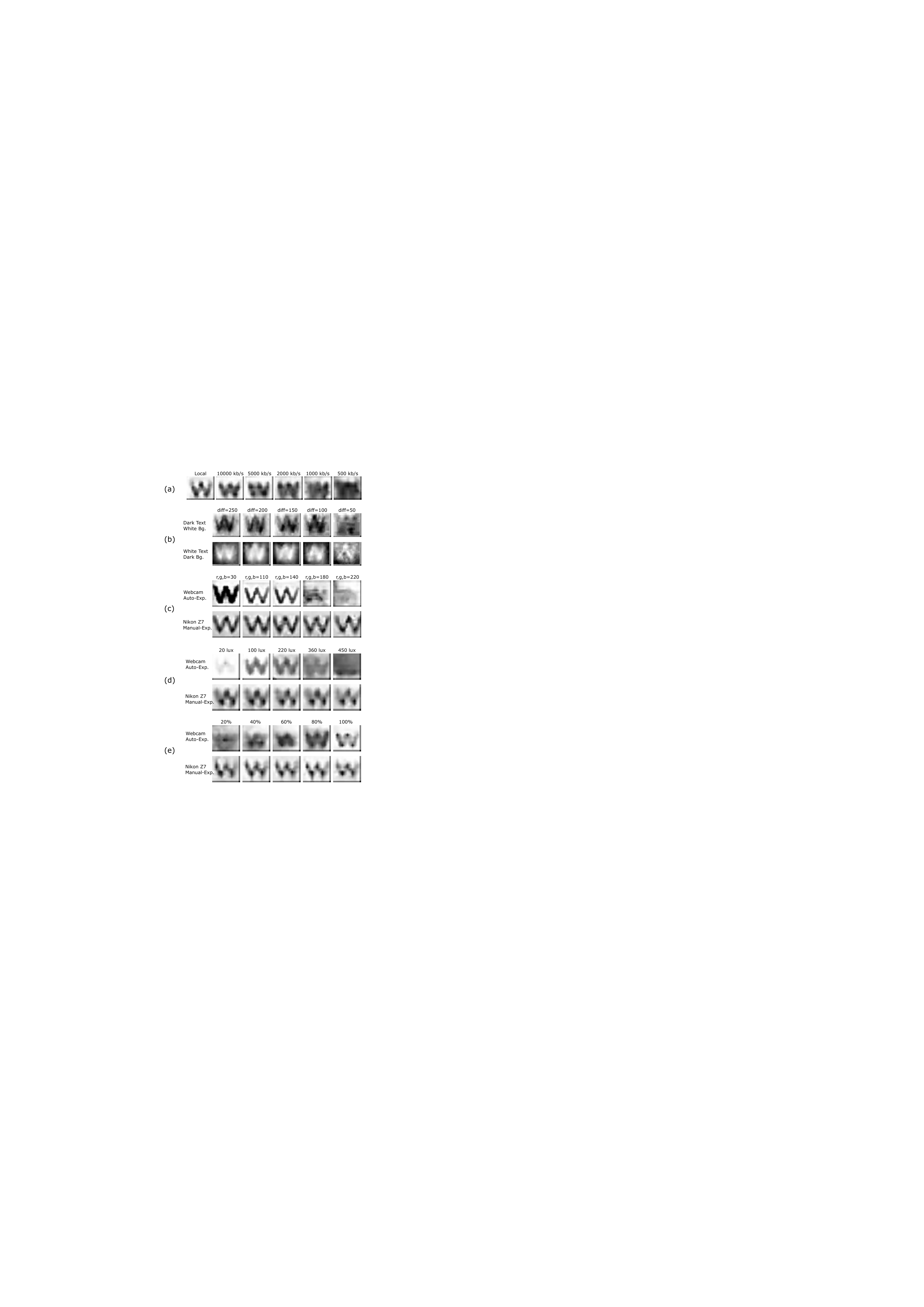}
    % \vspace{-.1in}
	\caption{(a) The comparison between reconstructed images when the video is recorded locally on the victim device and over Zoom with different network upload bandwidths. (b) Changes of reflection recognizability with different text-background color contrast. (c) Changes of reflection recognizability with different background colors (reflectance). We tested gray-scale colors with the same RGB values, which have relatively uniform reflectance on the visible light spectrum. (d) Changes of reflection recognizability under different environmental light intensities. (e) Changes in reflection recognizability with different screen brightness.} 
	\label{fig:factors_visual}
\end{figure}

% \section{User Study Factor Estimation} \label{apdx:est}
% The overall environmental light intensity and is estimated by taking the average pixel luminance in the video frames. The screen brightness is estimated by taking the average pixel luminance in glass area containing the screen reflections. The glass-screen distance is estimated by the physical average physiognomical face height \cite{farkas2005international} and physiognomical pixel height in the video frames using similar formula as Equation~\ref{eq:reflect}.  

\end{document}